\documentclass[abstract,headings=normal]{scrartcl}

\usepackage[automark]{scrpage2}
\usepackage[english]{babel}
\usepackage[T1]{fontenc}
\usepackage[applemac]{inputenc}
\usepackage{graphicx}
\usepackage{amsmath}
\usepackage{amssymb}
\usepackage{amsthm}
\usepackage{subfig}
\usepackage{hyperref}

\newcommand{\C}{\mathbb{C}}
\newcommand{\Z}{\mathbb{Z}}
\newcommand{\CP}{\mathbb{CP}}
\newcommand{\D}{\mathcal{D}}
\newcommand{\Pe}{\mathcal{P}}
\newcommand{\Dach}{\textasciicircum}

\DeclareMathOperator{\sn}{sn}

\newcommand{\Moeb}{\text{M\"ob}}

\newcommand{\Hvier}{H\textsuperscript{4}}
\newcommand{\Hsechs}{H\textsuperscript{6}}

\newtheorem{theo}{Theorem}[section]
\newtheorem{lemma}[theo]{Lemma}
\theoremstyle{definition}

\theoremstyle{remark}
\newtheorem*{rem}{Remark}

\graphicspath{{./graphics/}}
\DeclareGraphicsExtensions{.pdf}
  
\begin{document}

\title{Classification of 3D consistent quad-equations}

\author{Raphael Boll\footnote{Institut f\"ur Mathematik, MA 7-2, Technische Universit\"at Berlin, Str.~des~17.~Juni~136, 10623 Berlin, Germany; e-mail: boll@math.tu-berlin.de}}

\maketitle

\begin{abstract} We consider 3D consistent systems of six independent quad-equations assigned to the faces of a cube. The well-known classification of 3D consistent quad-equations, the so-called ABS-list, is included in this situation. The extension of these equations to the whole lattice $\Z^{3}$ is possible by reflecting the cubes. For every quad-equation we will give at least one system included leading to a B\"acklund transformation and a zero-curvature representation which means that they are integrable.\par
\vspace{1cm}
\end{abstract}

\section{Introduction}
One of the definitions of integrability of lattice equations, which becomes increasingly popular in the recent years, is based on the notion of multidimensional consistency. For two-dimensional lattices, this notion was clearly formulated first in \cite{NW}, and it was proposed to use as a synonym of integrability in \cite{quadgraphs,Nijhoff}. The outstanding importance of 3D consistency in the theory of discrete integrable systems became evident no later than with the appearance of the well-known ABS-classification of integrable equations in \cite{ABS1}. In that article Adler, Bobenko and Suris used a  definition of 3D consistency only allowing equations on the faces of a cube, which differ only by the parameter values assigned to the edges of a cube. In \cite{Atk1} and \cite{Todapaper} appeared a lot of systems of quad-equations not satisfying this strict definition of 3D consistency. However, these systems can also be seen as families of B\"acklund transformations and they lead to zero curvature representations of participating quad-equations in the same way (see \cite{quadgraphs}).\par
As already done in \cite{ABS2} the definition of 3D consistency can be extended: In contrast to the restriction, that all faces of a cube must carry the same equation up to parameters assigned to edges of the cube, we will allow different equations on all faces of a cube. The classification in that article is restricted to so-called equations of type~Q, i.e.\ those whose biquadratics are all non-degenerate (we will give a precise definition in the next section). The present paper is devoted to systems containing equations which are not necessarily of type Q. Our classification will cover systems appearing in \cite{Atk1} and all systems in \cite{Todapaper}, as well as equations which are equivalent to the Hietarinta equation \cite{Hietarinta}, to the “new” equation in \cite{HV} and to the equation in \cite{LY}. Moreover, it will contains also many novel systems.\par
In addition, every system in this classification can be extended to the whole lattice $\Z^{3}$ by reflecting the cubes.\par
The outline of our approach is the following: In Section~\ref{faces} we will present a complete classification of a single quad-equation modulo M\"obius transformations acting independently on the fields at the four vertices of an elementary quadrilateral. In Section~\ref{cubes} we will give a classification of 3D consistent systems of quad-equations possessing the so-called tetrahedron property modulo M\"obius transformations acting independently on the fields at the eight vertices of an elementary cube. In Section~\ref{embedding} we will show how to embed our systems in the lattice $\Z^{3}$ and how to derive B\"acklund transformations and zero curvature representations from our systems. This will include the idea of embedding considered in \cite{XP} as a special case and can be seen as a justification for the extended definition of 3D consistency to yield a definition of integrability.

\section{Quad-Equations on Single Quadrilaterals} \label{faces}
At the beginning we will introduce some objects and notations. We will start with the most important one, the \emph{quad-equation} $Q\left(x_{1},x_{2},x_{3},x_{4}\right)=0$, where $Q\in\C\left[x_{1},x_{2},x_{3},x_{4}\right]$ is an irreducible multi-affine polynomial.\par
Very useful tools for characterizing quad-equations, are the \emph{biquadratics}. We define them for every permutation $\left(i,j,k,\ell\right)$ of $\left(1,2,3,4\right)$ as follows
\[Q^{i,j}=Q^{i,j}\left(x_{i},x_{j}\right)=Q_{x_{k}}Q_{x_{\ell}}-QQ_{x_{k},x_{\ell}}.\]
A biquadratic $h\left(x,y\right)$ is called \emph{non-degenerate} if no polynomial in its equivalence class with respect to M\"obius transformations in $x$ and $y$ is divisible by a factor $x-\alpha_{1}$ or $y-\alpha_{2}$ (with $\alpha_{i}\in\C$). Otherwise, $h$ is called \emph{degenerate} and factors $x-\alpha_{1}$ and $y-\alpha_{2}$ with $\left(x-\alpha_{1}\right)\mid h$ and $\left(y-\alpha_{2}\right)\mid h$ are called \emph{factors of degeneracy}. Moreover,  if $x\mid x^{2}\cdot h\left(1/x,y\right)$, we write $\left(x-\infty\right)\mid h\left(x,y\right)$.\par
The Theorem 2 in \cite{ABS2} and earlier (in a different context) \cite{biquad} gives a complete classification of biquadratics up to M\"obius transformations. 
In particular, it can be shown that a biquadratic is degenerate if and only if $i_{3}=0$, where for a biquadratic $h\left(x,y\right)$ its \emph{relative invariant} $i_{3}$ is defined by
\[i_{3}\left(h,x,y\right)=\frac{1}{4}\det\begin{pmatrix}h&h_{x}&h_{xx}\\h_{y}&h_{xy}&h_{xxy}\\h_{yy}&h_{xyy}&h_{xxyy}\end{pmatrix}.\]
A multi-affine polynomial $Q$ is of \emph{type~Q} if all its biquadratics are non-degenerate. Otherwise it is of \emph{type~\Hvier} if four out of six biquadratics are degenerate and of \emph{type~\Hsechs} if all six biquadratics are degenerate. According to Lemmas~\ref{gegenueber} and \ref{nicht2} which we will prove later there are no other possibilities for $Q$.\par
For every permutation $\left(i,j,k,\ell\right)$ of $\left(1,2,3,4\right)$ the quartic polynomial
\[r^{i}=r^{i}\left(x_{i}\right)=\left(Q^{j,k}_{x_{\ell}}\right)^{2}-2Q^{j,k}Q^{j,k}_{x_{\ell},x_{\ell}}\]
is called a corresponding \emph{discriminant}. This polynomial turns out to be independent on permutations of $\left(j,k,\ell\right)$\par
Let $\Pe_{n}^{m}$ denote the set of polynomials in $n$ variables which are of degree $m$ in each variable. We consider the following action of M\"obius transformations on polynomials $f\in\Pe_{n}^{m}$:
\[M\left[f\right]\left(x_{1},\ldots,x_{n}\right)=\left(c_{1}x_{1}+d_{1}\right)^{m}\cdots\left(c_{n}x_{n}+d_{n}\right)^{m}f\left(\frac{a_{1}x_{1}+b_{1}}{c_{1}x_{1}+d_{1}},\ldots,\frac{a_{n}x_{n}+b_{n}}{c_{n}x_{n}+d_{n}}\right),\]
where $a_{i}d_{i}-b_{i}c_{i}\neq0$. The group $\left(\Moeb\right)^{4}$ acts on quad-equations by M\"obius transformations on all fields independently. \par
We will now present a complete classification of quad-equations on single quadrilaterals. We will not give the complete proofs here, because they are to long. However, in Section~\ref{tools} we give an overview of the most important ingredients of this proofs.

\subsection{Quad-Equations of Type Q}
Quad-equations of type Q were already classified in \cite{ABS2}. Every quad-equation of type Q is equivalent modulo $\left(\Moeb\right)^{4}$ to one of the following quad-equations characterized by the quadruples of discriminants:
\begin{itemize}
\item $\left(\delta, \delta, \delta, \delta\right)$:
\begin{equation}
Q=\alpha_{2}\left(x_{1}-x_{2}\right)\left(x_{3}-x_{4}\right)-\alpha_{1}\left(x_{1}-x_{4}\right)\left(x_{2}-x_{3}\right)+\delta\alpha_{1}\alpha_{2}\left(\alpha_{1}+\alpha_{2}\right)\tag{$Q_{1}$}\end{equation}
\item $\left(x_{1},x_{2},x_{3},x_{4}\right)$:
\begin{multline}
Q=\alpha_{2}\left(x_{1}-x_{2}\right)\left(x_{3}-x_{4}\right)-\alpha_{1}\left(x_{1}-x_{4}\right)\left(x_{2}-x_{3}\right)\\
+\alpha_{1}\alpha_{2}\left(\alpha_{1}+\alpha_{2}\right)\left(x_{1}+x_{2}+x_{3}+x_{4}\right)
-\alpha_{1}\alpha_{2}\left(\alpha_{1}+\alpha_{2}\right)\left(\alpha_{1}^{2}+\alpha_{1}\alpha_{2}+\alpha_{2}^{2}\right)\tag{$Q_{2}$}\end{multline}
\item $\left(x_{1}^{2}-\delta,x_{2}^{2}-\delta,x_{3}^{2}-\delta,x_{4}^{2}-\delta\right)$:
\begin{multline}
Q=\left(\alpha_{1}-\alpha_{1}^{-1}\right)\left(x_{1}x_{2}+x_{3}x_{4}\right)+\left(\alpha_{2}-\alpha_{2}^{-1}\right)\left(x_{1}x_{4}+x_{2}x_{3}\right)\\
 -\left(\alpha_{1}\alpha_{2}-\alpha_{1}^{-1}\alpha_{2}^{-1}\right)\left(x_{1}x_{3}+x_{2}x_{4}\right)+\frac{\delta}{4}\left(\alpha_{1}-\alpha_{1}^{-1}\right)\left(\alpha_{2}-\alpha_{2}^{-1}\right)\left(\alpha_{1}\alpha_{2}-\alpha_{1}^{-1}\alpha_{2}^{-1}\right)\tag{$Q_{3}$}\end{multline}
\item $\left(\left(x_{1}-1\right)\left(k^{2}x_{1}^{2}-1\right),\left(x_{2}-1\right)\left(k^{2}x_{2}^{2}-1\right),\left(x_{3}-1\right)\left(k^{2}x_{3}^{2}-1\right),\left(x_{4}-1\right)\left(k^{2}x_{4}^{2}-1\right)\right)$:
\begin{multline}
Q=\sn\left(\alpha_{1}\right)\sn\left(\alpha_{2}\right)\sn\left(\alpha_{1}+\alpha_{2}\right)\left(k^{2}x_{1}x_{2}x_{3}x_{4}+1\right)-\sn\left(\alpha_{1}\right)\left(x_{1}x_{2}+x_{3}x_{4}\right)\\
-\sn\left(\alpha_{2}\right)\left(x_{1}x_{4}+x_{2}x_{3}\right)+\sn\left(\alpha_{1}+\alpha_{2}\right)\left(x_{1}x_{3}+x_{2}x_{4}\right)\tag{$Q_{4}$}\end{multline}\end{itemize}

\subsection{Quad-Equations of Type \texorpdfstring{\Hvier}{H4} and \texorpdfstring{\Hsechs}{H6}}
A complete classification of type \Hvier\ and of type \Hsechs\ quad-equations did not appear in the literature before. Every quad-equation of type \Hvier\  is equivalent modulo $\left(\Moeb\right)^{4}$ to one of the following quad-equations characterized by the quadruples of discriminants:
\begin{itemize}
\item $\left(\epsilon,0,\epsilon,0\right)$:
\begin{equation}
Q=\left(x_{1}-x_{3}\right)\left(x_{2}-x_{4}\right)+\left(\alpha_{2}-\alpha_{1}\right)\left(1+\epsilon x_{2}x_{4}\right)\tag{$H_{1}^{\epsilon}$}\end{equation}
\item $\left(\epsilon x_{1}+1,1,\epsilon x_{3}+1,1\right)$:
\begin{multline}
Q=\left(x_{1}-x_{3}\right)\left(x_{2}-x_{4}\right)+\left(\alpha_{2}-\alpha_{1}\right)\left(x_{1}+x_{2}+x_{3}+x_{4}\right)+\alpha_{2}^{2}-\alpha_{1}^{2}\\
+\epsilon\left(\alpha_{2}-\alpha_{1}\right)\left(2 x_{2}+\alpha_{1}+\alpha_{2}\right)\left(2 x_{4}+\alpha_{1}+\alpha_{2}\right)+\epsilon\left(\alpha_{2}-\alpha_{1}\right)^{3}\tag{$H_{2}^{\epsilon}$}\end{multline}
\item $\left(x_{1}^{2}+\delta\epsilon,x_{2}^{2},x_{3}^{2}+\delta\epsilon,x_{4}^{2}\right)$:
\begin{equation}Q=\alpha_{1}\left(x_{1}x_{2}+x_{3}x_{4}\right)-\alpha_{2}\left(x_{1}x_{4}+x_{2}x_{3}\right)+\left(\alpha_{1}^{2}-\alpha_{2}^{2}\right)\left(\delta+\frac{\epsilon x_{2}x_{4}}{\alpha_{1}\alpha_{2}}\right)\tag{$H_{3}^{\epsilon}$}\end{equation}\end{itemize}
\begin{rem} All these equations were already mentioned in \cite{ABS2}.\end{rem}\par
Every quad-equation of type \Hsechs\  is equivalent modulo $\left(\Moeb\right)^{4}$ to one of the following quad-equations characterized by the quadruples of discriminants:
\begin{itemize}
\item $\left(0,0,0,0\right)$:
\[Q=x_{1}+x_{2}+x_{3}+x_{4}\]
\item $\left(1,0,1,\delta\right)$:
\[Q=x_{1}+x_{3}+x_{2}\left(x_{4}+\delta x_{1}\right)\]
\item $\left(x_{1}^{2},x_{2}^{2},x_{3}^{2},x_{4}^{2}\right)$:
\[Q=x_{1}x_{3}+x_{2}x_{4}+\delta_{1}x_{2}x_{3}+\delta_{2}x_{3}x_{4}\]\end{itemize}

\subsection{Ingredients of the proofs} \label{tools}
We will now present some ingredients of the proofs needed for the classification of quad-equations. At this point we will repeat two formulas already given in \cite{ABS2}:
\begin{equation} \label{hformel} 4i_{3}\left(Q^{1,2},x_{1},x_{2}\right)Q^{1,4}=\det\begin{pmatrix}Q^{1,2}&Q^{1,2}_{x_{1}}&\ell\\Q^{1,2}_{x_{2}}& Q^{1,2}_{x_{1}x_{2}}&\ell_{x_{2}}\\Q^{1,2}_{x_{2}x_{2}}&Q^{1,2}_{x_{1}x_{2}x_{2}}&\ell_{x_{2}x_{2}}\end{pmatrix},\end{equation}
where \[\ell=Q^{2,3}_{x_{3}x_{3}}Q^{3,4}-Q^{2,3}_{x_{3}}Q^{3,4}_{x_{3}}+Q^{2,3}Q^{3,4}_{x_{3}x_{3}}\] and
\begin{equation} \label{Qformel} \frac{2Q_{x_{1}}}{Q}=\frac{Q^{1,2}_{x_{1}}Q^{3,4}-Q^{1,4}_{x_{1}}Q^{2,3}+Q^{2,3}Q^{3,4}_{x_{3}}-Q^{2,3}_{x_{3}}Q^{3,4}}{Q^{1,2}Q^{3,4}-Q^{1,4}Q^{2,3}}.\end{equation}
The following Lemma gives some informations about the relation between non-de\-gen\-er\-ate biquadratics of a quad-equation: 
\begin{lemma} \label{gegenueber} Biquadratics on opposite edges (we consider the two diagonals as opposite edges, too) are either both degenerate or both non-degenerate.
\begin{proof} A biquadratic $Q^{i,j}$ is degenerate if and only if $i_{3}=0$ holds. Due to \cite{ABS2} $i_{3}$ is equal for biquadratics on opposite edges.\end{proof}\end{lemma}
Moreover, it follows that the number of non-degenerate biquadratics of a quad-e\-qua\-tion is even. Another restriction for biquadratics of a quad-equation comes along with the next Lemma:
\begin{lemma} \label{nicht2} There do not exist any quad-equation with exactly two degenerate biquadratics.
\begin{proof} \emph{Assumption:} $Q$ is such a quad-equation. Change variables in a way, that the two degenerate biquadratics are $Q^{1,3}$ and $Q^{2,4}$. Then, all biquadratics on edges are non-degenerate. According to \cite{ABS2} $Q$ is of Type Q and all biquadratics are non-degenerate. Contradiction!\end{proof}\end{lemma}
In addition, one can show, that $Q^{i,j}\not\equiv0$:
\begin{lemma} Every biquadratic of a quad-equation is not the zero polynomial.
\begin{proof}
By a simple calculation one can show
\begin{equation} \label{2biquads} Q\left(Q_{x_{2}}^{1,2}-Q_{x_{3}}^{1,3}\right)=2\left(Q_{x_{3}}Q^{1,3}-Q_{x_{2}}Q^{1,2}\right).\end{equation}
Let $Q^{1,2}=0$. Then, due to the classification of biquadratics we get $r^{1}=r^{2}= 0$. Assume that $Q^{1,3}\neq0$. We have to consider two cases:
\begin{itemize}
\item If $Q^{1,3}$ is non-degenerate, suitable M\"obius transformations in $x_{1}$ and $x_{3}$ lead to $Q^{1,3}=\left(x_{1}-x_{3}\right)^{2}$. In the same manner, we get $Q^{2,4}=\left(x_{2}-x_{4}\right)^{2}$, and using \eqref{hformel} we get $Q^{3,4}=0$. Now, one can apply \eqref{Qformel} and arrive $Q=\left(x_{1}-x_{3}\right)\left(x_{2}-x_{4}\right)$.
\item Otherwise, up to M\"obius transformation in $x_{3}$ we get $Q_{x_{3}}^{1,3}=0$ according to the classification of biquadratics. Using \eqref{2biquads} we get $Q_{x_{3}}=0$.\end{itemize}
Both cases are not possible because $Q$ is irreducible. Therefore, all biquadratics must be zero polynomials. Consider
\[\left(\log Q\right)_{x_{1}x_{2}}=\left(\frac{Q_{x_{1}}}{Q}\right)_{x_{2}}=\frac{QQ_{x_{1}x_{2}}-Q_{x_{1}}Q_{x_{2}}}{Q^{2}}=-\frac{h^{3,4}}{Q^{2}}=0\]
and in the same manner
\[\left(\log Q\right)_{x_{1}x_{3}}=\left(\log Q\right)_{x_{1}x_{4}}=\left(\log Q\right)_{x_{2}x_{3}}=\left(\log Q\right)_{x_{2}x_{4}}=\left(\log Q\right)_{x_{3}x_{4}}=0.\]
Therefore,
\[\log Q=\phi_{1}\left(x_{1}\right)+\phi_{2}\left(x_{1}\right)+\phi_{3}\left(x_{3}\right)+\phi_{4}\left(x_{4}\right)\]
and, furthermore,
\[Q=\phi_{1}\left(x_{1}\right)\phi_{2}\left(x_{1}\right)\phi_{3}\left(x_{3}\right)\phi_{4}\left(x_{4}\right)\]
with $\phi_{i}\left(x_{i}\right)=\alpha_{i}x_{i}+\beta_{i}$ which is reducible. Contradiction!\end{proof}\end{lemma}
Moreover, we also have the following lemma concerning vanishing biquadratics:
\begin{lemma} \label{110} There is no solution $\left(x_{1},x_{2},x_{3},x_{4}\right)$ of $Q\left(x_{1},x_{2},x_{3},x_{4}\right)=0$ with $Q^{1,2}\neq0$, $Q^{1,3}\neq0$ and $Q^{1,4}=0$.
\begin{proof}
Let $\left(x_{1},x_{2},x_{3},x_{4}\right)$ of $Q\left(x_{1},x_{2},x_{3},x_{4}\right)=0$ with $Q^{1,2}\neq0$, $Q^{1,3}\neq0$ and $Q^{1,4}=0$. Then, $Q^{1,2}\neq0$ leads to $Q_{x_{3}}\neq0$ and $Q^{1,3}\neq0$ leads to $Q_{x_{2}}\neq0$. This is a contradiction to $Q_{x_{2}}Q_{x_{3}}=0$ which is equivalent to $Q^{1,4}=0$.\end{proof}\end{lemma}
Now, we are able to proof the following lemma:
\begin{lemma} \label{facdeg} Every factor of degeneracy is a factor of at least two biquadratics, that means if $\left(x-\alpha\right)\mid h^{1,2}$ then
\[\left(x_{1}-\alpha\right)\mid h^{1,3}\ \text{or}\ \left(x_{1}-\alpha\right)\mid h^{1,4}.\]
\begin{proof}
We have to consider the following cases:
\begin{enumerate}
\item $\deg_{x_{1}}h^{1,4}\in\left\{0,1\right\}$\par
Considering the M\"obius transformation $x_{1}\mapsto x_{1}+\alpha$ we have to show: If $x_{1}\mid h^{1,2}$, then $x_{1}\mid h^{1,3}$ or $x_{1}\mid h^{1,4}$.\par
\emph{Assumption:} $x_{1}\mid h^{1,2}$ but $x_{1}\nmid h^{1,3}$ and $x_{1}\nmid h^{1,4}$.\par
We set $N_{i}:=\left\{x\in\CP^{1}:h^{1,i}\left(0,x\right)=0\right\}$ with $i\in\left\{3,4\right\}$. Obviously, $\left|N_{3}\right|,\left|N_{4}\right|<\infty$.
Due to Lemma \ref{110} there exists no solution $\left(x_{1},x_{2},x_{3},x_{4}\right)$ of $Q\left(x_{1},x_{2},x_{3},x_{4}\right)=0$ with $x_{1}=0$, $x_{3}\notin N_{3}$ and $x_{4}\notin N_{4}$.\par
$Q$ can be written as $Q=px_{1}+q$ with $p,q\in\C\left[x_{2},x_{3},x_{4}\right]$. $q_{x_{2}}=0$ (if not, $q=\tilde{p}x_{2}+\tilde{q}$ with $\tilde{p},\tilde{q}\in\C\left[x_{3},x_{4}\right]$ and $\tilde{p}\neq0$. Therefore, there would be $x_{3},x_{4}\in\C\setminus\left(N_{3}\cup N_{4}\right)$, such that $\tilde{p}\left(x_{3},x_{4}\right)\neq0$ and therefore, \[Q\left(0,-\frac{\tilde{q}\left(x_{3},x_{4}\right)}{\tilde{p}\left(x_{3},x_{4}\right)},x_{3},x_{4}\right)=0).\] In the same manner one can show, that $q_{x_{3}}=q_{x_{4}}=0$ and therefore without restriction  $q=1$.\par
Then, $h^{1,4}=p_{x_{2}}x_{1}p_{x_{3}}x_{1}-p_{x_{2},x_{3}}x_{1}\left(px_{1}+1\right)$. From $\deg_{x_{1}}h^{1,4}\leq1$ there follows, $p_{x_{2}}p_{x_{3}}-p_{x_{2},x_{3}}p=0$ and therefore $h^{1,4}=-p_{x_{2}x_{3}}x_{1}$. Contradiction!
\item $h^{1,4}=\left(x_{1}-\epsilon\right)^{2}X$ with $X\in\C\left[x_{4}\right]$\par
If $\epsilon\neq\alpha$, we reach the first case using the M\"obius transformation $x_{1}\mapsto \frac{\epsilon x_{1}+1}{x_{1}}$ which leads to $h^{1,4}=X$.
\item $h^{1,4}=\left(x_{1}-\epsilon\right)\left(x_{1}-\hat{\epsilon}\right)X$ with $X\in\C\left[x_{4}\right]$ and $\epsilon\neq\hat{\epsilon}$\par
If $\epsilon\neq\alpha$ and $\hat{\epsilon}\neq\alpha$, we reach the first case using the M\"obius transformation $x_{1}\mapsto \frac{\epsilon x_{1}+1}{x_{1}}$ or $x_{1}\mapsto \frac{\hat{\epsilon} x_{1}+1}{x_{1}}$ which leads to $h^{1,4}=\left(x_{1}\pm\frac{1}{\hat{\epsilon}-\epsilon}\right)X$.\end{enumerate}\end{proof}\end{lemma}
These Lemmas are the necessary tools for the classification of quad-equations.

\section{Quad-Equations on the Faces of a Cube} \label{cubes}
\begin{figure}[htbp]
   \centering
   \subfloat[Normal Case]{\label{fig:cube}\includegraphics{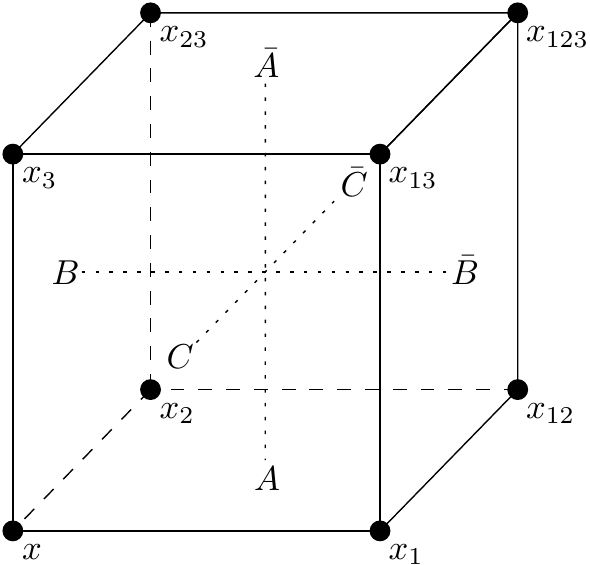}}\qquad
   \subfloat[Flipped Case]{\label{fig:cube2}\includegraphics{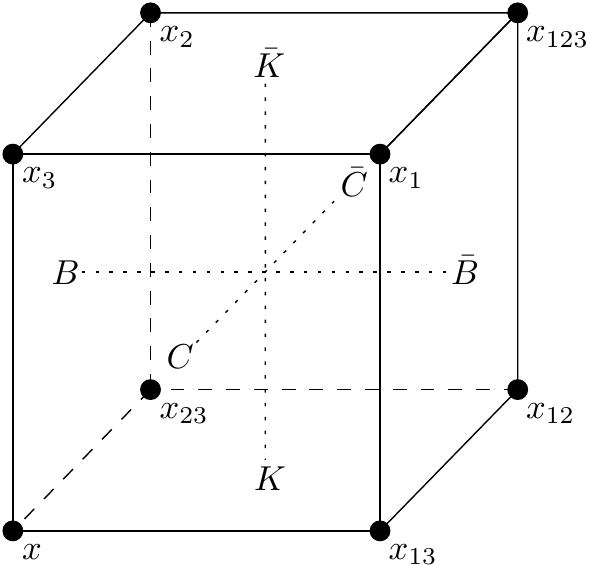}}
   \caption{Equations on a Cube}
\end{figure}
We will now consider systems of the type
\begin{align} \label{system} A\left(x,x_{1},x_{2},x_{12}\right)&=0,&
\bar{A}\left(x_{3},x_{13},x_{23},x_{123}\right)&=0,\notag\\
B\left(x,x_{2},x_{3},x_{23}\right)&=0,&
\bar{B}\left(x_{1},x_{12},x_{13},x_{123}\right)&=0,\\
C\left(x,x_{1},x_{3},x_{13}\right)&=0,&
\bar{C}\left(x_{2},x_{12},x_{23},x_{123}\right)&=0\notag\end{align}
where the equations $A,\ldots,\bar{C}$ are quad-equations assigned to the faces of a cube in the manner demonstrated in Figure~\ref{fig:cube}. Such a system is \emph{3D consistent} if the three values for $x_{123}$ (calculated by using $\bar{A}=0$, $\bar{B}=0$ or $\bar{C}=0$) coincide for arbitrary initial data $x$, $x_{1}$, $x_{2}$ and $x_{3}$. It possesses the \emph{tetrahedron property} if there exist two polynomials $K$ and $\bar{K}$ such that the equations
\begin{align*}
&K\left(x,x_{12},x_{13},x_{23}\right)=0& \text{ and}&
&\bar{K}\left(x_{1},x_{2},x_{3},x_{123}\right)=0\end{align*}
are satisfied for every solution of the system. It can be shown that the polynomials $K$ and $\bar{K}$ are multi-affine and irreducible. For this proof we use the following Lemma:
\begin{lemma} \label{3biquads}Consider a 3D consistent system \eqref{system} and
\begin{align*} F\left(x,x_{1},x_{2},x_{3},x_{123}\right)&=\bar{A}_{x_{13},x_{23}}BC-\bar{A}_{x_{23}}BC_{x_{13}}-\bar{A}_{x_{13}}B_{x_{23}}C+\bar{A}B_{x_{23}}C_{x_{13}},\\
 G\left(x,x_{1},x_{2},x_{3},x_{123}\right)&=\bar{B}_{x_{12},x_{13}}CA-\bar{B}_{x_{13}}CA_{x_{12}}-\bar{B}_{x_{12}}C_{x_{13}}A+\bar{B}C_{x_{13}}A_{x_{12}},\\
 H\left(x,x_{1},x_{2},x_{3},x_{123}\right)&=\bar{C}_{x_{12},x_{23}}AB-\bar{C}_{x_{12}}AB_{x_{23}}-\bar{C}_{x_{23}}A_{x_{12}}B+\bar{C}A_{x_{12}}B_{x_{23}}.\end{align*}
Then,
\[F=G=H=0\]
and
\[ F_{x_{1}}F_{x_{2}}-F_{x_{1},x_{2}}F=B^{0,3}C^{0,3}\bar{A}^{3,123}.\]
\begin{proof}
We get the equations
\[F=G=H=0\]
by eliminating $x_{12}$, $x_{13}$ and $x_{23}$ in the system \eqref{system}, and we have simply to factorize $F_{x_{1}}F_{x_{2}}-F_{x_{1},x_{2}}F$ to get the second statement.\end{proof}\end{lemma}
This allows us to proof the following Lemma:
\begin{lemma}\label{Kirreducible} Consider a 3D consistent system~\eqref{system} possessing the tetrahedron property described by the two equations
\begin{align*}
K\left(x,x_{12},x_{13},x_{23}\right)&=0\\
\bar{K}\left(x_{1},x_{2},x_{3},x_{123}\right)&=0.\end{align*}
Then, $K$ and $\bar{K}$ are multi-affine, irreducible polynomials.
\begin{proof}
Consider the system~\eqref{system}. The elimination of $x_{12}$, $x_{13}$ and $x_{23}$ leads to
\begin{align*} F\left(\overset{2}{x},\overset{1}{x}_{1},\overset{1}{x}_{2},\overset{3}{x}_{3},\overset{1}{x}_{123}\right)&=\bar{A}_{x_{13},x_{23}}BC-\bar{A}_{x_{23}}BC_{x_{13}}-\bar{A}_{x_{13}}B_{x_{23}}C+\bar{A}B_{x_{23}}C_{x_{13}}=0,\\
 G\left(\overset{2}{x},\overset{3}{x}_{1},\overset{1}{x}_{2},\overset{1}{x}_{3},\overset{1}{x}_{123}\right)&=\bar{B}_{x_{12},x_{13}}CA-\bar{B}_{x_{13}}CA_{x_{12}}-\bar{B}_{x_{12}}C_{x_{13}}A+\bar{B}C_{x_{13}}A_{x_{12}}=0,\\
 H\left(\overset{2}{x},\overset{1}{x}_{1},\overset{3}{x}_{2},\overset{1}{x}_{3},\overset{1}{x}_{123}\right)&=\bar{C}_{x_{12},x_{23}}AB-\bar{C}_{x_{12}}AB_{x_{23}}-\bar{C}_{x_{23}}A_{x_{12}}B+\bar{C}A_{x_{12}}B_{x_{23}}=0
\end{align*}
where the numbers over the arguments of $F$, $G$ and $H$ indicate their degrees in the corresponding variables. These degrees are in the projective sense, that is in agreement with the action of M\"obius transformations. Therefore the polynomials $F$, $G$ and $H$ must factorize as:
\begin{align*} F&=f\left(x,\overset{2}{x}_{3}\right)L,&
G&=g\left(x,\overset{2}{x}_{1}\right)L,&
H&=h\left(x,\overset{2}{x}_{2}\right)L,&
L=L\left(x,\overset{1}{x}_{1},\overset{1}{x}_{2},\overset{1}{x}_{3},\overset{1}{x}_{123}\right)\end{align*}
and therefore $L=k\left(x,x_{1},x_{2},x_{3},x_{123}\right)\bar{K}$. Therefore, $\bar{K}$ is multi-affine.\par
Assume, that $\bar{K}$ is reducible. Then, without restriction $\bar{K}=x_{1}d\left(x_{2},x_{3},x_{123}\right)$ or $\bar{K}=\left(x_{1}-x_{3}\right)d\left(x_{2},x_{123}\right)$ (otherwise, change the labeling and apply some M\"obius transformations). In both cases $\bar{K}^{3,123}=0$.\par
Then, due to Lemma \ref{3biquads}
\[0=f^{2}k^{2}\bar{K}^{3,123}=F_{x_{1}}F_{x_{2}}-F_{x_{1},x_{2}}F=B^{0,3}C^{0,3}\bar{A}^{3,123}\]
which is a contradiction to $B^{0,3},C^{0,3},\bar{A}^{3,123}\not\equiv0$. Analogously, we can proof the same for $K$.\end{proof}\end{lemma}
The group $\left(\Moeb\right)^{8}$ acts on such a system by M\"obius transformations on all vertex fields independently.\par
We classify all 3D consistent systems \eqref{system} with a tetrahedron property, whereas the classification of the case, that all quad-equations are of type Q, was already done in \cite{ABS2}. Note, that in many cases the tetrahedron property is a consequence of the other assumptions.\par
There are two essential ideas which allow for this classification. The first one was already used in \cite{ABS2} and deals with the coincidence of biquadratics assigned to an edge but belonging to different faces. We will adapt the results from \cite{ABS2} to our situations. The second one is completely new and it can be interpreted as flipping certain vertices of a cube. We will present three theorems, two devoted to the first idea, one to the second one.
\begin{theo} \label{coincide1}
Consider a 3D consistent system~\eqref{system} with $B^{0,3}$ and $C^{0,3}$ are non-degen\-erate and
\begin{itemize}
\item $\bar{A}^{3,123}$ is non-degenerate or
\item the discriminants of $B$ and $C$ corresponding to the vertices of $x$ and $x_{3}$ are not equal to zero.
\end{itemize}
Then:
\begin{enumerate}
\item \eqref{system} possesses the tetrahedron property.
\item For any edge of the cube, the two biquadratics corresponding to this edge coincide up to a constant factor.
\item The product of this factors around one vertex is equal to $-1$; for example
\[A^{0,1}B^{0,2}C^{0,3}+A^{0,2}B^{0,3}C^{0,1}=0.\]
\end{enumerate}
\begin{proof}
The elimination of $x_{12}$, $x_{13}$ and $x_{23}$ leads to
\begin{align*} F\left(\overset{2}{x},\overset{1}{x}_{1},\overset{1}{x}_{2},\overset{3}{x}_{3},\overset{1}{x}_{123}\right)&=\bar{A}_{x_{13},x_{23}}BC-\bar{A}_{x_{23}}BC_{x_{13}}-\bar{A}_{x_{13}}B_{x_{23}}C+\bar{A}B_{x_{23}}C_{x_{13}}=0,\\
 G\left(\overset{2}{x},\overset{3}{x}_{1},\overset{1}{x}_{2},\overset{1}{x}_{3},\overset{1}{x}_{123}\right)&=\bar{B}_{x_{12},x_{13}}CA-\bar{B}_{x_{13}}CA_{x_{12}}-\bar{B}_{x_{12}}C_{x_{13}}A+\bar{B}C_{x_{13}}A_{x_{12}}=0,\\
 H\left(\overset{2}{x},\overset{1}{x}_{1},\overset{3}{x}_{2},\overset{1}{x}_{3},\overset{1}{x}_{123}\right)&=\bar{C}_{x_{12},x_{23}}AB-\bar{C}_{x_{12}}AB_{x_{23}}-\bar{C}_{x_{23}}A_{x_{12}}B+\bar{C}A_{x_{12}}B_{x_{23}}=0
\end{align*}
where the numbers over the arguments of $F$, $G$ and $H$ indicate their degrees in the corresponding variables.
Therefore the polynomials $F$, $G$ and $H$ must factorize as:
\begin{align*} F&=f\left(x,\overset{2}{x}_{3}\right)L,&
G&=g\left(x,\overset{2}{x}_{1}\right)L,&
H&=h\left(x,\overset{2}{x}_{2}\right)L,&
L=L\left(x,\overset{1}{x}_{1},\overset{1}{x}_{2},\overset{1}{x}_{3},\overset{1}{x}_{123}\right).\end{align*}
Then, due to Lemma \ref{3biquads}
\[f^{2}\left(L_{x_{1}}L_{x_{2}}-L_{x_{1},x_{2}}L\right)=F_{x_{1}}F_{x_{2}}-F_{x_{1},x_{2}}F=B^{0,3}C^{0,3}\bar{A}^{3,123}.\]
Consider first the case that $\bar{A}^{3,123}$ is non-degenerate. Then, $\bar{A}^{3,123}\mid\left(L_{x_{1}}L_{x_{2}}-L_{x_{1},x_{2}}L\right)$. Since $B^{0,3}$ and $C^{0,3}$ are non-degenerate, too, $B^{0,3}=f=C^{0,3}$ up to constant factors. In the other case $B^{0,3}$ and $C^{0,3}$ are not complete squares. Therefore, since $B^{0,3}$ and $C^{0,3}$ are non-degenerate $B^{0,3}C^{0,3}\mid f^{2}$ or $B^{0,3}C^{0,3}\mid\left(L_{x_{1}}L_{x_{2}}-L_{x_{1},x_{2}}L\right)$ which is not possible because of $\deg_{x}L=1$. Therefore, $B^{0,3}=f=C^{0,3}$ up to constant factors. Then, $\deg_{x}f=2$ and therefore $\deg_{x}L=0$, so the tetrahedron property is valid. According to \cite{ABS1} this is equivalent to
\[A^{0,1}B^{0,2}C^{0,3}+A^{0,2}B^{0,3}C^{0,1}=0.\]
Therefore, $A^{0,1}/C^{0,1}$ can only depend on $x$ and not on $x_{1}$.
Since for symmetry reasons also
\[A^{0,1}\bar{B}^{1,12}C^{1,13}+A^{1,12}\bar{B}^{1,13}C^{0,1}=0\]
holds, $A^{0,1}/C^{0,1}$ is constant. This completes the proof.\end{proof}\end{theo}
\begin{theo} \label{coincide2}
Consider a 3D consistent system~\eqref{system} with
\begin{itemize}
\item all discriminants on diagonals of faces are non-degenerate and
\item all discriminants not equal to zero.
\end{itemize}
Then:
\begin{enumerate}
\item For any edge of the cube, the two biquadratic polynomials corresponding to this edge coincide up to a constant factor.
\item If in addition system \eqref{system} possesses the tetrahedron property, the product of this factors around one vertex is equal to $-1$; for example,
\[A^{0,1}B^{0,2}C^{0,3}+A^{0,2}B^{0,3}C^{0,1}=0.\]
\end{enumerate}
\begin{proof}
The elimination of $x_{12}$, $x_{13}$ and $x_{23}$ leads to
\begin{align*} F\left(\overset{2}{x},\overset{1}{x}_{1},\overset{1}{x}_{2},\overset{3}{x}_{3},\overset{1}{x}_{123}\right)&=\bar{A}_{x_{13},x_{23}}BC-\bar{A}_{x_{23}}BC_{x_{13}}-\bar{A}_{x_{13}}B_{x_{23}}C+\bar{A}B_{x_{23}}C_{x_{13}}=0,\\
 G\left(\overset{2}{x},\overset{3}{x}_{1},\overset{1}{x}_{2},\overset{1}{x}_{3},\overset{1}{x}_{123}\right)&=\bar{B}_{x_{12},x_{13}}CA-\bar{B}_{x_{13}}CA_{x_{12}}-\bar{B}_{x_{12}}C_{x_{13}}A+\bar{B}C_{x_{13}}A_{x_{12}}=0,\\
 H\left(\overset{2}{x},\overset{1}{x}_{1},\overset{3}{x}_{2},\overset{1}{x}_{3},\overset{1}{x}_{123}\right)&=\bar{C}_{x_{12},x_{23}}AB-\bar{C}_{x_{12}}AB_{x_{23}}-\bar{C}_{x_{23}}A_{x_{12}}B+\bar{C}A_{x_{12}}B_{x_{23}}=0
\end{align*}
where the numbers over the arguments of $F$, $G$ and $H$ indicate their degrees in the corresponding variables.
Therefore the polynomials $F$, $G$ and $H$ must factorize as:
\begin{align*} F&=f\left(x,\overset{2}{x}_{3}\right)L,&
G&=g\left(x,\overset{2}{x}_{1}\right)L,&
H&=h\left(x,\overset{2}{x}_{2}\right)L,&
L=L\left(x,\overset{1}{x}_{1},\overset{1}{x}_{2},\overset{1}{x}_{3},\overset{1}{x}_{123}\right).\end{align*}
Then, due to Lemma \ref{3biquads}
\[f^{2}\left(L_{x_{1}}L_{x_{2}}-L_{x_{1},x_{2}}L\right)=F_{x_{1}}F_{x_{2}}-F_{x_{1},x_{2}}F=B^{0,3}C^{0,3}\bar{A}^{3,123}\]
and since $\bar{A}^{3,123}$ is non-degenerate and $B^{0,3}$ and $C^{0,3}$ are not of Type $\left(0,0\right)$, i.e. $B^{0,3}$ and $C^{0,3}$ are not complete squares, we have $B^{0,3}=f\cdot k_{1}\left(x\right)$ and $C^{0,3}=f\cdot k_{2}\left(x\right)$ with some polynomials $k_{1}$ and $k_{2}$, so that $B^{0,3}/C^{0,3}$ can depend on $x$ only.
Analogously, the elimination of $x_{1}$, $x_{2}$ and $x_{123}$ leads to
\[\bar{f}^{2}\left(\bar{L}_{x_{13}}\bar{L}_{x_{23}}-\bar{L}_{x_{13},x_{23}}\bar{L}\right)=B^{0,3}C^{0,3}A^{0,12}.\]
Since $A^{0,12}$ is non-degenerate, too, we have $B^{0,3}=\bar{f}\cdot \bar{k}_{1}\left(x_{3}\right)$ and $C^{0,3}=\bar{f}\cdot \bar{k}_{2}\left(x_{3}\right)$ with some polynomials $\bar{k}_{1}$ and $\bar{k}_{2}$, so that $B^{0,3}/C^{0,3}$ can depend on $x_{3}$ only. Therefore, $B^{0,3}=\hat{k}C^{0,3}$ with $\hat{k}\in\C$.\par
In \cite{ABS1} it is shown, that the tetrahedron property is equivalent to
\[A^{0,1}B^{0,2}C^{0,3}+A^{0,2}B^{0,3}C^{0,1}=0.\]
This completes the proof.\end{proof}\end{theo}
\begin{theo} \label{tetrahedronUse} Consider a 3D consistent system~\eqref{system} possessing the tetrahedron property described by the two equations
\begin{align*}
K\left(x,x_{12},x_{13},x_{23}\right)&=0\\
\bar{K}\left(x_{1},x_{2},x_{3},x_{123}\right)&=0.\end{align*}
Then, the system
\begin{align}\label{dualsystem} K\left(x,x_{12},x_{13},x_{23}\right)&=0,&
\bar{K}\left(x_{1},x_{2},x_{3},x_{123}\right)&=0,\notag\\
B\left(x,x_{2},x_{3},x_{23}\right)&=0,&
\bar{B}\left(x_{1},x_{12},x_{13},x_{123}\right)&=0,\\
C\left(x,x_{1},x_{3},x_{13}\right)&=0,&
\bar{C}\left(x_{2},x_{12},x_{23},x_{123}\right)&=0,\notag\end{align}
 which can be assigned to a cube in the manner demonstrated in Figure~\ref{fig:cube2} on page~\pageref{fig:cube2},
is 3D consistent and possesses the tetrahedron property. 3D consistency of \eqref{dualsystem} is understood as the property of the initial value problem with initial date $x$, $x_{3}$, $x_{13}$ and $x_{23}$.
\begin{proof}
Let $x$, $x_{3}$, $x_{13}$ and $x_{23}$ be the initial data for the system~\eqref{dualsystem}. Then, we can calculate $x_{1}$ using $C\left(x,x_{1},x_{3},x_{13}\right)=0$, $x_{2}$ using $B\left(x,x_{2},x_{3},x_{23}\right)=0$ and $x_{12}$ using $K\left(x,x_{12},x_{13},x_{23}\right)=0$.\par
Furthermore, one can proof that $A\left(x,x_{1},x_{2},x_{12}\right)=0$ for this values of $x_{1}$, $x_{2}$ and $x_{12}$: Assume that $A\left(x,x_{1},x_{2},x_{12}\right)=0$ is not satisfied and let $x_{1}$ and $x_{2}$ be fixed. Then, we would get another value $\bar{x}_{12}$ for $x_{12}$ using $A\left(x,x_{1},x_{2},x_{12}\right)=0$. Since \eqref{system} possesses the tetrahedron property, $K\left(x,\bar{x}_{12},x_{13},x_{23}\right)=0$  would hold, but due to Lemma~\ref{Kirreducible} this is a contradiction to $K\left(x,x_{12},x_{13},x_{23}\right)=0$.\par
With this values of $x_{1}$, $x_{2}$ and $x_{12}$ use the equations $\bar{B}\left(x_{1},x_{12},x_{13},x_{123}\right)=0$ and $\bar{C}\left(x_{2},x_{12},x_{23},x_{123}\right)=0$ to calculate two values for $x_{123}$ which are equal because of the 3D consistency of \eqref{system}. Moreover, $\bar{A}\left(x_{3},x_{13},x_{23},x_{123}\right)=0$ is satisfied due to the 3D consistency of \eqref{system} and $\bar{K}\left(x_{1},x_{2},x_{3},x_{123}\right)=0$ is satisfied, because \eqref{system} possesses the tetrahedron property.\end{proof}\end{theo}
We will now present the classification. The proofs will only be given for the first two sections. The other proofs are quite analogous.

\subsection{Six Equations of Type \texorpdfstring{\Hvier}{H4}, First Case}
In this section we consider systems~(\ref{system}) with
\begin{itemize}
\item $A,\bar{A},\ldots,\bar{C}$ of type \Hvier\  and
\item all non-degenerate biquadratics on diagonals of faces.
\end{itemize}
Below is the list of all 3D consistent systems modulo $\left(\Moeb\right)^{8}$ with these properties and with the tetrahedron property. It turns out that their tetrahedron property follows from the above assumptions except for the system characterized by the quadruple $\left(\epsilon,0,0,\epsilon\right)$.
\begin{theo} \label{Hvierfirst} Every 3D consistent system~(\ref{system}) satisfying the properties of this section and possessing the tetrahedron property is equivalent modulo $\left(\Moeb\right)^{8}$ to one of the following three systems. They are written in terms of two polynomials $A\left(x,x_{1},x_{2},x_{12};\alpha,\beta\right)$ and $K\left(x,x_{12},x_{13},x_{23};\epsilon\right)$ as
\begin{align*}
 \bar{A}&=A\left(x_{13},x_{3},x_{123},x_{23};\alpha,\beta\right),&
B&=A\left(x,x_{2},x_{3},x_{23};\beta,\gamma\right),\\
 \bar{B}&=A\left(x_{12},x_{1},x_{123},x_{13};\beta,\gamma\right),&
C&=A\left(x,x_{1},x_{3},x_{13};\alpha,\gamma\right),\\
 \bar{C}&=A\left(x_{12},x_{2},x_{123},x_{23};\alpha,\gamma\right),&
 \bar{K}&=K\left(x_{1},x_{2},x_{3},x_{123};0\right).\end{align*}
The polynumials $A$ and $K$ can be characterized by the quadruples of discriminants of $A$:
\begin{itemize}
\item $\left(\epsilon,0,0,\epsilon\right)$:
\begin{align*}
A=&\left(x-x_{12}\right)\left(x_{1}-x_{2}\right)-\left(\alpha-\beta\right)\left(1+\epsilon x_{1}x_{2}\right),\\
\begin{split}K=&\left(\beta-\gamma\right)\left(x-x_{12}\right)\left(x_{13}-x_{23}\right)-\left(\alpha-\beta\right)\left(x-x_{23}\right)\left(x_{12}-x_{13}\right)\\
&-\epsilon\left(\alpha-\beta\right)\left(\beta-\gamma\right)\left(\alpha-\gamma\right)\end{split}\end{align*}
\item $\left(\epsilon x+1,1,1,\epsilon x_{12}+1\right)$:
\begin{align*}
 \begin{split}A=&\left(x-x_{12}\right)\left(x_{1}-x_{2}\right)-\left(\alpha-\beta\right)\left(x+x_{1}+x_{2}+x_{12}\right)-\alpha^{2}+\beta^{2}\\
 &-\epsilon\left(\alpha-\beta\right)\left(2x_{1}+\alpha+\beta\right)\left(2x_{2}+\alpha+\beta\right)-\epsilon\left(\alpha-\beta\right)^{3},\end{split}\\
\begin{split} K=&\left(\beta-\gamma\right)\left(x-x_{12}\right)\left(x_{13}-x_{23}\right)-\left(\alpha-\beta\right)\left(x-x_{23}\right)\left(x_{12}-x_{13}\right)\\
&+\left(\alpha-\beta\right)\left(\beta-\gamma\right)\left(\alpha-\gamma\right)-2\epsilon\left(\alpha-\beta\right)\left(\beta-\gamma\right)\left(\alpha-\gamma\right)\left(x+x_{12}+x_{13}+x_{23}\right)\\
&-4\epsilon^{2}\left(\alpha-\beta\right)\left(\beta-\gamma\right)\left(\alpha-\gamma\right)\left(\left(\alpha-\beta\right)^{2}+\left(\alpha-\beta\right)\left(\beta-\gamma\right)+\left(\beta-\gamma\right)^{2}\right)\end{split}\end{align*}
\item $\left(x^{2}+\delta\epsilon,x_{1}^{2},x_{2}^{2},x_{12}^{2}+\delta\epsilon\right)$:
\begin{align*}
A=&\alpha\left(xx_{1}+x_{2}x_{12}\right)-\beta\left(xx_{2}+x_{1}x_{12}\right)
+\left(\alpha^{2}-\beta^{2}\right)\left(\delta+\frac{\epsilon x_{1}x_{2}}{\alpha\beta}\right),\\
\begin{split}K=&\gamma\left(\alpha^{2}-\beta^{2}\right)\left(xx_{12}+x_{13}x_{23}\right)-\beta\left(\alpha^{2}-\gamma^{2}\right)\left(xx_{13}+x_{12}x_{23}\right)\\
&+\alpha\left(\beta^{2}-\gamma^{2}\right)\left(xx_{23}+x_{12}x_{13}\right)+\frac{\delta\epsilon\left(\alpha^{2}-\beta^{2}\right)\left(\alpha^{2}-\gamma^{2}\right)\left(\beta^{2}-\gamma^{2}\right)}{\alpha\beta\gamma}\end{split}\end{align*}\end{itemize}
\begin{proof}
We will start with systems characterized by $\left(\epsilon,0,0,\epsilon\right)$. In this case we suppose that the tetrahedron property holds. Due Section~\ref{faces} we have
\[A=\left(x-x_{12}\right)\left(x_{1}-x_{2}\right)-\alpha\left(1+\epsilon x_{1}x_{2}\right)\]
up to M\"obius transformations in $x$, $x_{1}$, $x_{2}$ and $x_{12}$. We have the biquadratics
\begin{align*}
&A^{0,1}=\alpha\left(1+\epsilon x_{1}^{2}\right),&
&A^{0,2}=-\alpha\left(1+\epsilon x_{2}^{2}\right),&
&A^{2,12}=\alpha\left(1+\epsilon x_{2}^{2}\right).\end{align*}
The biquadratics $A^{0,2}$ and $B^{0,2}$ coincide up to a constant factor because of the tetrahedron property. Therefore, up to M\"obius transformations in $x_{3}$ and $x_{23}$ we have
\[B=\left(x-x_{23}\right)\left(x_{2}-x_{3}\right)-\beta\left(1+\epsilon x_{2}x_{3}\right)\]
with biquadratics
\begin{align*}
&B^{0,2}=\beta\left(1+\epsilon x_{2}^{2}\right),&
&B^{0,3}=-\beta\left(1+\epsilon x_{3}^{2}\right),&
&B^{2,23}=-\beta\left(1+\epsilon x_{2}^{2}\right).\end{align*}
We have to keep in mind that M\"obius transformations $x\mapsto \mu x+\nu$ and, if $\epsilon=0$, also $x_{2}\mapsto \mu_{2} x+\nu_{2}$ do not change $B^{0,2}$ up to a constant factor. However, the influence of these transformations on $B$ can be eliminated by $x_{23}\mapsto \mu x_{23}+\nu$ and $\beta\mapsto\mu\beta$ or, if $\epsilon=0$, by $x_{3}\mapsto \mu_{2}x_{3}+\nu_{2}$, $x_{23}\mapsto \mu x_{23}+\nu$ and $\beta\mapsto\mu\mu_{2}\beta$.
Furthermore, the biquadratics $A^{0,1}$ and $C^{0,1}$ as well as $B^{0,3}$ and $C^{0,3}$ coincide up to a constant factor and, moreover, we have
\[A^{0,1}B^{0,2}C^{0,3}+A^{0,2}B^{0,3}C^{0,1}=0.\]
Therefore, up to M\"obius transformation in $x_{13}$ we have
\[C=\left(x-x_{13}\right)\left(x_{1}-x_{3}\right)-\gamma\left(1+\epsilon x_{1}x_{3}\right)+\tilde{\gamma}\left(x-x_{13}\right)\]
with $\tilde{\gamma}=0$, if $\epsilon\neq0$, and biquadratics
\begin{align*}
&C^{0,1}=\gamma\left(1+\epsilon x_{1}^{2}\right),&
&C^{0,3}=\gamma\left(1+\epsilon x_{3}^{2}\right).\end{align*}
Again, we have to keep in mind that M\"obius transformations $x\mapsto \mu x+\nu$ and, if $\epsilon=0$, also $x_{1}\mapsto \mu_{1}x_{1}+\nu_{1}$ and $x_{3}\mapsto \mu_{1}x_{3}+\nu_{3}$ do not change $C^{0,1}$ and $C^{0,3}$ up to a common constant factor. However, the influence of these transformations on $C$ can be eliminated by $x_{13}\mapsto\mu x_{13}+\nu$ and $\gamma\mapsto\mu\gamma$ or, if $\epsilon=0$, $x_{13}\mapsto\mu x_{13}+\nu$, $\gamma\mapsto\mu\mu_{1}\gamma$ and $\tilde{\gamma}\mapsto \mu_{1}\tilde{\gamma}-\nu_{1}+\nu_{3}$.\par
From $A$, $B$ and $C$ one can derive $K$. However, $K$ is multi-affine and independent on $x_{1}$, $x_{2}$ and $x_{3}$ only if $\gamma=\alpha+\beta$ and $\tilde{\gamma}=0$ hold. We get
\[K=\alpha\left(x-x_{23}\right)\left(x_{12}-x_{13}\right)-\beta\left(x-x_{12}\right)\left(x_{13}-x_{23}\right)+\epsilon\alpha\beta\left(\alpha+\beta\right)\]
with the biquadratic
\[K^{12,23}=\alpha\beta\left(\left(x_{12}-x_{23}\right)^{2}+\epsilon\left(\alpha+\beta\right)^{2}\right).\]
In the same way as for $C$, we get, up to M\"obius transformation in $x_{123}$,
\[\bar{C}=\left(x_{2}-x_{123}\right)\left(x_{12}-x_{23}\right)-\gamma_{2}\left(1+\epsilon x_{12}x_{23}\right)+\tilde{\gamma}_{2}\left(x_{2}-x_{123}\right)\] with the biquadratic
\[\bar{C}^{12,23}=-\left(\left(x_{12}-x_{23}+\bar{\gamma}_{2}\right)^{2}+\epsilon\gamma_{2}^{2}\right).\]
Due to Theorem~\ref{tetrahedronUse} we have $\gamma_{2}=\alpha+\beta$ and $\bar{\gamma}_{2}=0$. $\bar{A}$, $\bar{B}$ and $\bar{K}$ can now easily derived from the other equations. After transformations $\alpha\mapsto\alpha-\beta$ and $\beta\mapsto\beta-\gamma$ we get the above system.\par
Now, we consider the systems characterized by $\left(\epsilon x+1,1,1,\epsilon x_{12}+1\right)$. In this case we do not suppose the tetrahedron property. We will show, that it follows from the above assumptions. Due to Section~\ref{faces} we have
\begin{multline*}B=\left(x-x_{23}\right)\left(x_{2}-x_{3}\right)-\left(\beta-\gamma\right)\left(x+x_{2}+x_{3}+x_{23}\right)-\beta^{2}+\gamma^{2}-\\\epsilon\left(\beta-\gamma\right)\left(2x_{2}+\beta+\gamma\right)\left(2x_{3}+\beta+\gamma\right)-\epsilon\left(\beta-\gamma\right)^{3}\end{multline*}
up to M\"obius transformations in $x$, $x_{2}$, $x_{3}$ and $x_{23}$. We have the following biquadratics
\begin{align*}
B^{0,2}&=2\left(\beta-\gamma\right)\left(x+x_{2}+\beta+2\epsilon\left(x_{2}+\beta\right)^{2}\right),\\
B^{3,23}&=2\left(\beta-\gamma\right)\left(x_{3}+x_{23}+\beta+2\epsilon\left(x_{3}+\beta\right)^{2}\right),\\
B^{0,3}&=-2\left(\beta-\gamma\right)\left(x+x_{3}+\gamma+2\epsilon\left(x_{3}+\gamma\right)^{2}\right),\\
B^{2,23}&=-2\left(\beta-\gamma\right)\left(x_{2}+x_{23}+\gamma+2\epsilon\left(x_{2}+\gamma\right)^{2}\right).\end{align*}
Due to Theorem~\ref{coincide2} the biquadratics $A^{0,2}$ and $B^{0,2}$ coincide up to a constant factor. Therefore, we have, up to M\"obius transformations in $x_{1}$ and $x_{12}$,
\begin{multline*}
A=\left(x-x_{12}\right)\left(x_{1}-x_{2}\right)-\left(\alpha-\beta\right)\left(x+x_{1}+x_{2}+x_{12}\right)-\alpha^{2}+\beta^{2}-\\\epsilon\left(\alpha-\beta\right)\left(2x_{1}+\beta+\alpha\right)\left(2x_{2}+\beta+\alpha\right)-\epsilon\left(\alpha-\beta\right)^{3}\end{multline*}
with biquadratics
\begin{align*}
A^{0,1}&=2\left(\alpha-\beta\right)\left(x+x_{1}+\alpha+2\epsilon\left(x_{1}+\alpha\right)^{2}\right),\\
A^{2,12}&=2\left(\alpha-\beta\right)\left(x_{2}+x_{12}+\alpha+2\epsilon\left(x_{2}+\alpha\right)^{2}\right),\\
A^{0,2}&=-2\left(\alpha-\beta\right)\left(x+x_{2}+\beta+2\epsilon\left(x_{2}+\beta\right)^{2}\right),\\
A^{1,12}&=-2\left(\alpha-\beta\right)\left(x_{1}+x_{12}+\beta+2\epsilon\left(x_{1}+\beta\right)^{2}\right).\end{align*}
Furthermore, the biquadratics $A^{0,1}$ and $C^{0,1}$ as well as $B^{0,3}$ and $C^{0,3}$ coincide up to a constant factor, and therefore, we have, up to M\"obius transformation in $x_{13}$,
\begin{multline*}C=\left(x-x_{13}\right)\left(x_{3}-x_{1}\right)-\left(\gamma-\alpha\right)\left(x+x_{1}+x_{3}+x_{13}\right)-\gamma^{2}+\alpha^{2}-\\\epsilon\left(\gamma-\alpha\right)\left(2x_{1}+\gamma+\alpha\right)\left(2x_{3}+\gamma+\alpha\right)-\epsilon\left(\gamma-\alpha\right)^{3}\end{multline*}
with biquadratics
\begin{align*}
C^{0,1}&=-2\left(\gamma-\alpha\right)\left(x+x_{1}+\alpha+2\epsilon\left(x_{1}+\alpha\right)^{2}\right),\\
C^{3,13}&=-2\left(\gamma-\alpha\right)\left(x_{3}+x_{13}+\alpha+2\epsilon\left(x_{3}+\alpha\right)^{2}\right),\\
C^{0,3}&=2\left(\gamma-\alpha\right)\left(x+x_{3}+\gamma+2\epsilon\left(x_{3}+\gamma\right)^{2}\right),\\
C^{1,13}&=2\left(\gamma-\alpha\right)\left(x_{1}+x_{13}+\gamma+2\epsilon\left(x_{1}+\gamma\right)^{2}\right).\end{align*}
Moreover, the biquadratics $A^{2,12}$ and $\bar{C}^{2,12}$ as well as $B^{2,23}$ and $\bar{C}^{2,23}$ coincide up to a constant factor. Therefore, we have, up to M\"obius transformation in $x_{123}$,
\begin{multline*}\bar{C}=\left(x_{2}-x_{123}\right)\left(x_{23}-x_{12}\right)-\left(\gamma-\alpha\right)\left(x_{2}+x_{12}+x_{23}+x_{123}\right)-\gamma^{2}+\alpha^{2}-\\\epsilon\left(\gamma-\alpha\right)\left(2x_{2}+\gamma+\alpha\right)\left(2x_{123}+\gamma+\alpha\right)-\epsilon\left(\gamma-\alpha\right)^{3}\end{multline*}
with biquadratics
\begin{align*}
\bar{C}^{2,12}&=-2\left(\gamma-\alpha\right)\left(x_{2}+x_{12}+\alpha+2\epsilon\left(x_{2}+\alpha\right)^{2}\right),\\
\bar{C}^{23,123}&=-2\left(\gamma-\alpha\right)\left(x_{23}+x_{123}+\alpha+2\epsilon\left(x_{123}+\alpha\right)^{2}\right),\\
\bar{C}^{2,23}&=2\left(\gamma-\alpha\right)\left(x_{2}+x_{23}+\gamma+2\epsilon\left(x_{2}+\gamma\right)^{2}\right),\\
\bar{C}^{12,123}&=2\left(\gamma-\alpha\right)\left(x_{12}+x_{123}+\gamma+2\epsilon\left(x_{123}+\gamma\right)^{2}\right).\end{align*}
In addition, the biquadratics $\bar{A}^{3,13}$ and $C^{3,13}$, $\bar{A}^{3,23}$ and $B^{3,23}$ as well as $\bar{A}^{23,123}$ and $\bar{C}^{23,123}$ coincide up to a constant factor and therefore, we have
\begin{multline*}
\bar{A}=\left(x_{3}-x_{123}\right)\left(x_{13}-x_{23}\right)-\left(\alpha-\beta\right)\left(x_{3}+x_{13}+x_{23}+x_{123}\right)-\alpha^{2}+\beta^{2}-\\\epsilon\left(\alpha-\beta\right)\left(2x_{3}+\beta+\alpha\right)\left(2x_{123}+\beta+\alpha\right)-\epsilon\left(\alpha-\beta\right)^{3}\end{multline*}
with biquadratics
\begin{align*}
\bar{A}^{3,13}&=2\left(\alpha-\beta\right)\left(x_{3}+x_{13}+\alpha+2\epsilon\left(x_{3}+\alpha\right)^{2}\right),\\
\bar{A}^{23,123}&=2\left(\alpha-\beta\right)\left(x_{23}+x_{123}+\alpha+2\epsilon\left(x_{123}+\alpha\right)^{2}\right),\\
\bar{A}^{3,23}&=-2\left(\alpha-\beta\right)\left(x_{3}+x_{23}+\beta+2\epsilon\left(x_{3}+\beta\right)^{2}\right),\\
\bar{A}^{13,123}&=-2\left(\alpha-\beta\right)\left(x_{13}+x_{123}+\beta+2\epsilon\left(x_{123}+\beta\right)^{2}\right).\end{align*}
Nevertheless, $A^{1,12}$ and $\bar{B}^{1,12}$, $\bar{B}^{1,13}$ and $C^{1,13}$, $\bar{A}^{13,123}$ and $\bar{B}^{13,123}$ as well as $\bar{B}^{12,123}$ and $\bar{C}^{12,123}$ coincide up to a constant factor. Therefore, we have
\begin{multline*}\bar{B}=\left(x_{1}-x_{123}\right)\left(x_{12}-x_{13}\right)-\left(\beta-\gamma\right)\left(x_{1}+x_{12}+x_{13}+x_{123}\right)-\beta^{2}+\gamma^{2}-\\\epsilon\left(\beta-\gamma\right)\left(2x_{1}+\beta+\gamma\right)\left(2x_{123}+\beta+\gamma\right)-\epsilon\left(\beta-\gamma\right)^{3}\end{multline*}
with biquadratics
\begin{align*}
\bar{B}^{1,12}&=2\left(\beta-\gamma\right)\left(x_{1}+x_{12}+\beta+2\epsilon\left(x_{1}+\beta\right)^{2}\right),\\
\bar{B}^{13,123}&=2\left(\beta-\gamma\right)\left(x_{13}+x_{123}+\beta+2\epsilon\left(x_{123}+\beta\right)^{2}\right),\\
\bar{B}^{1,13}&=-2\left(\beta-\gamma\right)\left(x_{1}+x_{13}+\gamma+2\epsilon\left(x_{1}+\gamma\right)^{2}\right),\\
\bar{B}^{12,123}&=-2\left(\beta-\gamma\right)\left(x_{12}+x_{123}+\gamma+2\epsilon\left(x_{123}+\gamma\right)^{2}\right).\end{align*}
Now, one can easily compute $K$ and $\bar{K}$ from the above equations. Therefore, the tetrahedron property holds for this system.\par
The next case will consider on systems characterized by $\left(x^{2},x_{1}^{2},x_{2}^{2},x_{12}^{2}\right)$ where all biquadratics on edges have four factors of degeneracy. In this case we suppose the tetrahedron property. Due to Section~\ref{faces}
\[B=\beta\left(xx_{2}+x_{3}x_{23}\right)-\left(xx_{3}+x_{2}x_{23}\right)\]
up to M\"obius transformations in $x$, $x_{2}$, $x_{3}$ and $x_{23}$. We have the following biquadratics:
\begin{align*}
B^{0,2}&=-\left(\beta^{2}-1\right)xx_{2},&
B^{0,3}&=\left(\beta^{2}-1\right)xx_{3},&
B^{2,23}&=\left(\beta^{2}-1\right)x_{2}x_{23}.\end{align*}
The biquadratics $A^{0,2}$ and $B^{0,2}$ coincide up to a constant factor because of the tetrahedron property. Therefore, up to M\"obius transformations in $x_{1}$ and $x_{12}$ we have
\[A=\alpha\left(xx_{1}+x_{2}x_{12}\right)-\left(xx_{2}+x_{1}x_{12}\right)\]
with biquadratics
\begin{align*}
A^{0,1}&=-\left(\alpha^{2}-1\right)xx_{1},&
A^{0,2}&=\left(\alpha^{2}-1\right)xx_{2},&
A^{2,12}&=-\left(\alpha^{2}-1\right)x_{2}x_{12}.\end{align*}
We have to keep in mind that M\"obius transformations $x\mapsto x^{-1}$ and $x_{2}\mapsto x_{2}^{-1}$ do not change $A^{0,2}$ up to a constant factor. However, the influence of these transformations on $A$ can be eliminated by $x_{1}\mapsto x_{1}^{-1}$ and $x_{12}^{-1}$.
Furthermore, the biquadratics $A^{0,1}$ and $C^{0,1}$ as well as $B^{0,3}$ and $C^{0,3}$ coincide up to a constant factor and, moreover, we have
\[A^{0,1}B^{0,2}C^{0,3}+A^{0,2}B^{0,3}C^{0,1}=0.\]
Therefore, up to M\"obius transformation in $x_{13}$ we have
\[C=\gamma\left(xx_{3}+x_{1}x_{13}\right)-\left(xx_{1}+x_{3}x_{13}\right)\]
with biquadratics
\begin{align*}
C^{0,1}&=-\left(\gamma^{2}-1\right)xx_{1},&
C^{0,3}&=\left(\gamma^{2}-1\right)xx_{3}.\end{align*}
Again, we have to keep in mind that M\"obius transformations $x\mapsto x^{-1}$ as well as simultaneously $x_{1}\mapsto x_{1}^{\pm1}$ ans $x_{3}\mapsto x_{3}^{\pm1}$ do not change $C^{0,1}$ and $C^{0,3}$ up to a common constant factor. However, the influence of these transformations on $C$ can be eliminated by $x_{13}\mapsto x_{13}^{-1}$ as well as $\gamma\mapsto\gamma{\mp1}$.\par
From $A$, $B$ and $C$ one can derive $K$. However, $K$ is multi-affine and independent on $x_{1}$, $x_{2}$ and $x_{3}$ only if $\gamma=1/\left(\alpha\beta\right)$ hold. We get
\[K=\beta\left(\alpha^{2}-1\right)\left(xx_{12}+x_{13}x_{23}\right)+\alpha\left(\beta^{2}-1\right)\left(xx_{23}+x_{12}x_{13}\right)-\left(\alpha^{2}\beta^{2}-1\right)\left(xx_{13}+x_{12}x_{23}\right)\]
with the biquadratic
\[K^{12,23}=\left(\alpha^{2}-1\right)\left(\beta^{2}-1\right)\left(\alpha\beta x_{12}-x_{23}\right)\left(x_{12}-\alpha\beta x_{23}\right)\]
In the same way as for $C$, we get, up to M\"obius transformation in $x_{123}$,
\[\bar{C}=\gamma_{2}\left(x_{2}x_{23}+x_{12}x_{123}\right)-\left(x_{2}x_{12}+x_{23}x_{123}\right)\]
with the biquadratic
\[\bar{C}^{12,23}=-\left(\gamma_{2} x_{12}-x_{23}\right)\left(x_{12}-\gamma_{2} x_{23}\right).\]
Due Theorem~\ref{tetrahedronUse} we have $\gamma_{2}=\alpha\beta$. $\bar{A}$, $\bar{B}$ and $\bar{K}$ can now easily derived from the other equations. After transformations $\alpha\mapsto\alpha/\beta$ and $\beta\mapsto\beta/\gamma$ we get the above system with $\delta=\epsilon=0$.\par
Our last case will consider systems characterized by $\left(x^{2}+\delta\epsilon,x_{1}^{2},x_{2}^{2},x_{12}^{2}+\delta\epsilon\right)$ where all biquadratics have at most two factors of degeneracy. In this case we will also not suppose the tetrahedron property because it follows from the above assumptions. Due to Section~\ref{faces}
\[B=\beta\left(xx_{2}+x_{3}x_{23}\right)-\gamma\left(xx_{3}+x_{2}x_{23}\right)+\left(\beta^{2}-\gamma^{2}\right)\left(\delta+\frac{\epsilon x_{2}x_{3}}{\beta\gamma}\right)\]
up to M\"obius transformations in $x$, $x_{2}$, $x_{3}$ and $x_{23}$ with $\delta\neq0$ or $\epsilon\neq0$. We have the following biquadratics:
\begin{align*}
B^{0,2}&=-\left(\beta^{2}-\gamma^{2}\right)\left(xx_{2}+\delta\beta+\frac{\epsilon x_{2}^{2}}{\beta}\right),&
B^{0,3}&=\left(\beta^{2}-\gamma^{2}\right)\left(xx_{3}+\delta\gamma+\frac{\epsilon x_{3}^{2}}{\gamma}\right),\\
B^{3,23}&=-\left(\beta^{2}-\gamma^{2}\right)\left(x_{3}x_{23}+\delta\beta+\frac{\epsilon x_{3}^{2}}{\beta}\right),&
B^{2,23}&=\left(\beta^{2}-\gamma^{2}\right)\left(x_{2}x_{23}+\delta\gamma+\frac{\epsilon x_{2}^{2}}{\gamma}\right).\end{align*}
Due to Theorem~\ref{coincide2} the biquadratics $A^{0,2}$ and $B^{0,2}$ coincide up to a constant factor. Therefore, we have, up to M\"obius transformations in $x_{1}$ and $x_{12}$,
\[A=\alpha\left(xx_{1}+x_{2}x_{12}\right)-\beta\left(xx_{2}+x_{1}x_{12}\right)+\left(\alpha^{2}-\beta^{2}\right)\left(\delta+\frac{\epsilon x_{1}x_{2}}{\alpha\beta}\right)\]
with biquadratics
\begin{align*}
A^{0,1}&=-\left(\alpha^{2}-\beta^{2}\right)\left(xx_{1}+\delta\alpha+\frac{\epsilon x_{1}^{2}}{\alpha}\right),&
A^{0,2}&=\left(\alpha^{2}-\beta^{2}\right)\left(xx_{2}+\delta\beta+\frac{\epsilon x_{2}^{2}}{\beta}\right),\\
A^{2,12}&=-\left(\alpha^{2}-\beta^{2}\right)\left(x_{2}x_{12}+\delta\alpha+\frac{\epsilon x_{2}^{2}}{\alpha}\right),&
A^{1,12}&=\left(\alpha^{2}-\beta^{2}\right)\left(x_{1}x_{12}+\delta\beta+\frac{\epsilon x_{1}^{2}}{\beta}\right).\end{align*}
Furthermore, the biquadratics $A^{0,1}$ and $C^{0,1}$ as well as $B^{0,3}$ and $C^{0,3}$ coincide up to a constant factor, and therefore, we have, up to M\"obius transformation in $x_{13}$,
\[C=\gamma\left(xx_{3}+x_{1}x_{13}\right)-\alpha\left(xx_{1}+x_{3}x_{13}\right)+\left(\gamma^{2}-\alpha^{2}\right)\left(\delta+\frac{\epsilon x_{1}x_{3}}{\alpha\gamma}\right)\]
with biquadratics
\begin{align*}
C^{0,1}&=-\left(\alpha^{2}-\gamma^{2}\right)\left(xx_{1}+\delta\alpha+\frac{\epsilon x_{1}^{2}}{\alpha}\right),&
C^{0,3}&=\left(\alpha^{2}-\gamma^{2}\right)\left(xx_{3}+\delta\gamma+\frac{\epsilon x_{3}^{2}}{\gamma}\right),\\
C^{3,13}&=-\left(\alpha^{2}-\gamma^{2}\right)\left(x_{3}x_{13}+\delta\alpha+\frac{\epsilon x_{3}^{2}}{\alpha}\right),&
C^{1,13}&=\left(\alpha^{2}-\gamma^{2}\right)\left(x_{1}x_{13}+\delta\gamma+\frac{\epsilon x_{1}^{2}}{\gamma}\right).\end{align*}
Moreover, the biquadratics $A^{2,12}$ and $\bar{C}^{2,12}$ as well as $B^{2,23}$ and $\bar{C}^{2,23}$ coincide up to a constant factor. Therefore, we have, up to M\"obius transformation in $x_{123}$,
\[\bar{C}=\gamma\left(x_{2}x_{23}+x_{12}x_{123}\right)-\alpha\left(x_{2}x_{12}+x_{23}x_{123}\right)+\left(\gamma^{2}-\alpha^{2}\right)\left(\delta+\frac{\epsilon x_{2}x_{123}}{\alpha\gamma}\right)\]
with biquadratics
\begin{align*}
\bar{C}^{2,12}&=-\left(\alpha^{2}-\gamma^{2}\right)\left(x_{2}x_{12}+\delta\alpha+\frac{\epsilon x_{2}^{2}}{\alpha}\right),\\
\bar{C}^{23,123}&=-\left(\alpha^{2}-\gamma^{2}\right)\left(x_{23}x_{123}+\delta\alpha+\frac{\epsilon x_{123}^{2}}{\alpha}\right),\\
\bar{C}^{2,23}&=\left(\alpha^{2}-\gamma^{2}\right)\left(x_{2}x_{23}+\delta\gamma+\frac{\epsilon x_{2}^{2}}{\gamma}\right),\\
\bar{C}^{12,123}&=\left(\alpha^{2}-\gamma^{2}\right)\left(x_{12}x_{123}+\delta\gamma+\frac{\epsilon x_{123}^{2}}{\gamma}\right).\end{align*}
In addition, the biquadratics $\bar{A}^{3,13}$ and $C^{3,13}$, $\bar{A}^{3,23}$ and $B^{3,23}$ as well as $\bar{A}^{23,123}$ and $\bar{C}^{23,123}$ coincide up to a constant factor and therefore, we have
\[\bar{A}=\alpha\left(x_{3}x_{13}+x_{23}x_{123}\right)-\beta\left(x_{3}x_{23}+x_{13}x_{123}\right)+\left(\alpha^{2}-\beta^{2}\right)\left(\delta+\frac{\epsilon x_{3}x_{123}}{\alpha\beta}\right)\]
with biquadratics
\begin{align*}
\bar{A}^{3,13}&=-\left(\alpha^{2}-\beta^{2}\right)\left(x_{3}x_{13}+\delta\alpha+\frac{\epsilon x_{3}^{2}}{\alpha}\right),\\
\bar{A}^{23,123}&=-\left(\alpha^{2}-\beta^{2}\right)\left(x_{23}x_{123}+\delta\alpha+\frac{\epsilon x_{123}^{2}}{\alpha}\right),\\
\bar{A}^{3,23}&=\left(\alpha^{2}-\beta^{2}\right)\left(x_{3}x_{23}+\delta\beta+\frac{\epsilon x_{3}^{2}}{\beta}\right),\\
\bar{A}^{13,123}&=\left(\alpha^{2}-\beta^{2}\right)\left(x_{13}x_{123}+\delta\beta+\frac{\epsilon x_{123}^{2}}{\beta}\right).\end{align*}
Nevertheless, $A^{1,12}$ and $\bar{B}^{1,12}$, $\bar{B}^{1,13}$ and $C^{1,13}$, $\bar{A}^{13,123}$ and $\bar{B}^{13,123}$ as well as $\bar{B}^{12,123}$ and $\bar{C}^{12,123}$ coincide up to a constant factor. Therefore, we have
\[\bar{B}=\beta\left(x_{1}x_{12}+x_{13}x_{123}\right)-\gamma\left(x_{1}x_{13}+x_{12}x_{123}\right)+\left(\beta^{2}-\gamma^{2}\right)\left(\delta+\frac{\epsilon x_{1}x_{123}}{\beta\gamma}\right)\]
with biquadratics
\begin{align*}
\bar{B}^{1,12}&=-\left(\beta^{2}-\gamma^{2}\right)\left(x_{1}x_{12}+\delta\beta+\frac{\epsilon x_{1}^{2}}{\beta}\right),\\
\bar{B}^{13,123}&=-\left(\beta^{2}-\gamma^{2}\right)\left(x_{13}x_{123}+\delta\beta+\frac{\epsilon x_{123}^{2}}{\beta}\right),\\
\bar{B}^{1,13}&=\left(\beta^{2}-\gamma^{2}\right)\left(x_{1}x_{13}+\delta\gamma+\frac{\epsilon x_{1}^{2}}{\gamma}\right),\\
\bar{B}^{12,123}&=\left(\beta^{2}-\gamma^{2}\right)\left(x_{12}x_{123}+\delta\gamma+\frac{\epsilon x_{123}^{2}}{\gamma}\right).\end{align*}
Now, one can easily compute $K$ and $\bar{K}$ from the above equations. Therefore, the tetrahedron property holds for this system.\end{proof}\end{theo}
\begin{rem}
Equivalent systems appeared in \cite{ABS1,ABS2} without classification results for such systems.\end{rem}

\subsection{Two Equations of Type Q and Four Equations of Type \texorpdfstring{\Hvier}{H4}}
In this section we consider systems~(\ref{system}) with
\begin{itemize}
\item $B$ and $\bar{B}$ of type Q,
\item $A$, $\bar{A}$, $C$ and $\bar{C}$ of type \Hvier\  and
\item the non-degenerate biquadratics of $\bar{A}$, $C$ and $\bar{C}$ on edges neighboring $B$ or $\bar{B}$.
\end{itemize}
Below is the list of all 3D consistent systems modulo $\left(\Moeb\right)^{8}$ with these properties. It turns out that their tetrahedron property follows from the above assumptions.
\begin{theo}
Every 3D consistent system~(\ref{system}) satisfying the properties of this section is equivalent modulo $\left(\Moeb\right)^{8}$ to one of the following three systems. They are written in terms of the two polynomials $A\left(x,x_{1},x_{2},x_{12};\alpha,\beta\right)$ and $B\left(x,x_{2},x_{3},x_{23};\epsilon\right)$ as
\begin{align*}
 \bar{A}&=A\left(x_{3},x_{13},x_{23},x_{123};\alpha,\beta\right),&
 \bar{B}&=B\left(x_{1},x_{12},x_{13},x_{123};0\right),\\
 C&=A\left(x,x_{1},x_{3},x_{13};\alpha,\gamma\right),&
 \bar{C}&=A\left(x_{2},x_{12},x_{23},x_{123};\alpha,\gamma\right),\\
 K&=A\left(x,x_{12},x_{23},x_{13};\beta,\gamma\right),&
 \bar{K}&=A\left(x_{2},x_{1},x_{3},x_{123};\beta,\gamma\right).\end{align*}
The polynomials $A$ and $K$ can be characterized by the quadruples of discriminants of $A$:
\begin{itemize}
\item $\left(\epsilon,0,\epsilon,0\right)$:
\begin{align*}
 A=&\left(x-x_{2}\right)\left(x_{1}-x_{12}\right)-\left(\alpha-\beta\right)\left(1+\epsilon x_{1}x_{12}\right),\\
\begin{split}B=&\left(\beta-\gamma\right)\left(x-x_{2}\right)\left(x_{3}-x_{23}\right)-\left(\alpha-\beta\right)\left(x-x_{23}\right)\left(x_{2}-x_{3}\right)\\
&-\epsilon\left(\alpha-\beta\right)\left(\beta-\gamma\right)\left(\alpha-\gamma\right)\end{split}\end{align*}
\item $\left(\epsilon x+1,1,\epsilon x_{2}+1,1\right)$:
\begin{align*}
 \begin{split}A=&\left(x-x_{2}\right)\left(x_{1}-x_{12}\right)-\left(\alpha-\beta\right)\left(x+x_{1}+x_{2}+x_{12}\right)\\
&-\alpha^{2}+\beta^{2}-\epsilon\left(\alpha-\beta\right)\left(2x_{1}+\alpha+\beta\right)\left(2x_{12}+\alpha+\beta\right)-\epsilon\left(\alpha-\beta\right)^{3},\end{split}\\
\begin{split}B=&\left(\beta-\gamma\right)\left(x-x_{2}\right)\left(x_{3}-x_{23}\right)-\left(\alpha-\beta\right)\left(x-x_{23}\right)\left(x_{2}-x_{3}\right)\\
&+\left(\alpha-\beta\right)\left(\beta-\gamma\right)\left(\alpha-\gamma\right)-2\epsilon\left(\alpha-\beta\right)\left(\beta-\gamma\right)\left(\alpha-\gamma\right)\left(x+x_{2}+x_{3}+x_{23}\right)\\
&-4\epsilon^{2}\left(\alpha-\beta\right)\left(\beta-\gamma\right)\left(\alpha-\gamma\right)\left(\left(\alpha-\beta\right)^{2}+\left(\alpha-\beta\right)\left(\beta-\gamma\right)+\left(\beta-\gamma\right)^{2}\right)\end{split}\end{align*}
\item $\left(x^{2}+\delta\epsilon,x_{1}^{2},x_{2}^{2}+\delta\epsilon,x_{12}^{2}\right)$:
\begin{align*}
A=&\alpha\left(xx_{1}+x_{2}x_{12}\right)-\beta\left(xx_{12}+x_{1}x_{2}\right)+\left(\alpha^{2}-\beta^{2}\right)\left(\delta+\frac{\epsilon x_{1}x_{12}}{\alpha\beta}\right),\\
\begin{split}B=&\gamma\left(\alpha^{2}-\beta^{2}\right)\left(xx_{2}+x_{3}x_{23}\right)-\beta\left(\alpha^{2}-\gamma^{2}\right)\left(xx_{3}+x_{2}x_{23}\right)\\
&+\alpha\left(\beta^{2}-\gamma^{2}\right)\left(xx_{23}+x_{2}x_{3}\right)+\frac{\delta\epsilon\left(\alpha^{2}-\beta^{2}\right)\left(\alpha^{2}-\gamma^{2}\right)\left(\beta^{2}-\gamma^{2}\right)}{\alpha\beta\gamma}\end{split}\end{align*}\end{itemize}
\begin{proof}
 Due to Theorem~\ref{coincide1} all systems we have to consider possesses the tetrahedron property. Therefore, using Theorem~\ref{tetrahedronUse} we get our systems by the flips $x_{2}\leftrightarrow x_{12}$ and $x_{3}\leftrightarrow x_{13}$ in the systems of Theorem~\ref{Hvierfirst}.\end{proof}\end{theo}
\begin{rem}
Equivalent systems appeared in \cite{ABS2,Atk1,Todapaper} without classification results for such systems.\end{rem}

\subsection{Six Equations of Type \texorpdfstring{\Hvier}{H4}, Second Case}
In this section we consider systems~(\ref{system}) with
\begin{itemize}
\item $A,\bar{A},\ldots,\bar{C}$ of type \Hvier,
\item the non-degenerate biquadratics of $A$ and $\bar{A}$ on diagonals of faces and
\item the non-degenerate biquadratics of $B$, $\bar{B}$, $C$ and $\bar{C}$ on edges not neighboring $A$ and $\bar{A}$.
\end{itemize}
Below is the list of all 3D consistent systems modulo $\left(\Moeb\right)^{8}$ with these properties. It turns out that their tetrahedron property follows from the above assumptions.
\begin{theo} \label{Hviersecond} Every 3D consistent system~(\ref{system}) satisfying the properties of this section is equivalent modulo $\left(\Moeb\right)^{8}$ to one of the following three systems. They are written in terms of three polynomials $A\left(x,x_{1},x_{2},x_{12}\right)$, $B\left(x,x_{2},x_{3},x_{23};\beta\right)$ and $K\left(x,x_{12},x_{13},x_{23}\right)$ as
\begin{align*}
 \bar{A}&=A\left(x_{3},x_{13},x_{23},x_{123}\right),&
 \bar{B}&=B\left(x_{12},x_{1},x_{123},x_{13};\beta\right),&
 C&=B\left(x,x_{1},x_{3},x_{13};\alpha\right),\\
 \bar{C}&=B\left(x_{12},x_{2},x_{123},x_{23};\alpha\right),&
\bar{K}&=K\left(x_{3},x_{123},x_{1},x_{2}\right).\end{align*}
The polynomials $A$, $B$ and $K$ can be characterized by the quadruples of discriminants of $A$:
\begin{itemize}
\item $\left(\epsilon,0,0,\epsilon\right)$:
\begin{align*}
 &A=\left(x-x_{12}\right)\left(x_{1}-x_{2}\right)-\left(\alpha-\beta\right)\left(1+\epsilon x_{1}x_{2}\right),\\
&B=\left(x-x_{3}\right)\left(x_{2}-x_{23}\right)+\gamma\left(1+\epsilon x_{2}x_{23}\right),\\
&K=\left(x-x_{12}\right)\left(x_{13}-x_{23}\right)-\left(\alpha-\beta\right)\left(1+\epsilon x_{13}x_{23}\right)\end{align*}
\item $\left(\epsilon x+1,1,1,\epsilon x_{12}+1\right)$:
\begin{align*}
 \begin{split}A=&\left(x-x_{12}\right)\left(x_{1}-x_{2}\right)-\left(\alpha-\beta\right)\left(x+x_{1}+x_{2}+x_{12}\right)-\alpha^{2}+\beta^{2}\\
&-\epsilon\left(\alpha-\beta\right)\left(2 x_{1}+\alpha+\beta\right)\left(2 x_{2}+\alpha+\beta\right)-\epsilon\left(\alpha-\beta\right)^{3},\end{split}\\
 \begin{split}B=&\left(x-x_{3}\right)\left(x_{2}-x_{23}\right)+\gamma\left(x+x_{2}+x_{3}+x_{23}\right)+\gamma^{2}+2\beta\gamma\\
&+\epsilon \gamma\left(2 x_{2}+2\beta+\gamma\right)\left(2 x_{23}+2\beta+\gamma\right)+\gamma^{3}\epsilon,\end{split}\\
\begin{split}K=&\left(x-x_{12}\right)\left(x_{13}-x_{23}\right)-\left(\alpha-\beta\right)\left(x+x_{12}+x_{13}+x_{23}\right)-\alpha^{2}\beta^{2}\\
&-2\gamma\left(\alpha-\beta\right)-\epsilon\left(\alpha-\beta\right)\left(2x_{13}+\alpha+\beta+2\gamma\right)\left(2x_{23}+\alpha+\beta+2\gamma\right)-\epsilon\left(\alpha-\beta\right)^{3}\end{split}\end{align*}
\item $\left(x^{2}+\delta\epsilon,x_{1}^{2},x_{2}^{2},x_{12}^{2}+\delta\epsilon\right)$:
\begin{align*}
&A=\alpha\left(xx_{1}+x_{2}x_{12}\right)-\beta\left(xx_{2}+x_{1}x_{12}\right)+\left(\alpha^{2}-\beta^{2}\right)\left(\delta+\frac{\epsilon x_{1}x_{2}}{\alpha\beta}\right),\\
\begin{split}B=xx_{2}+x_{3}x_{23}-\gamma\left(xx_{23}+x_{2}x_{3}\right)-\left(\gamma^{2}-1\right)\left(\delta \beta +\frac{\epsilon x_{2}x_{23}}{\beta\gamma}\right),\end{split}\\
&K=\alpha\left(xx_{13}+x_{12}x_{23}\right)-\beta\left(xx_{23}+x_{12}x_{13}\right)+\left(\alpha^{2}-\beta^{2}\right)\left(\delta\gamma+\frac{\epsilon x_{13}x_{23}}{\alpha\beta\gamma}\right)\end{align*}\end{itemize}\end{theo}
\begin{rem}
These systems did not appear in the literature before.\end{rem}

\subsection{Four Equations of Type \texorpdfstring{\Hvier}{H4}\  and Two Equations of Type \texorpdfstring{\Hsechs}{H6}, First Case}
\begin{theo}No 3D consistent system~(\ref{system}) exists with the tetrahedron property and with
\begin{itemize}
\item $A$ and $\bar{A}$ of type \Hsechs,
\item $B$, $\bar{B}$, $C$ and $\bar{C}$ of type \Hvier,
\item the non-degenerate biquadratics of $B$, $\bar{B}$, $C$ and $\bar{C}$ on diagonals.
\end{itemize}\end{theo}

\subsection{Four Equations of Type \texorpdfstring{\Hvier}{H4}\  and Two Equations of Type \texorpdfstring{\Hsechs}{H6}, Second Case}
In this section we consider systems~(\ref{system}) with
\begin{itemize}
\item $A$ and $\bar{A}$ of type \Hsechs,
\item $B$, $\bar{B}$, $C$ and $\bar{C}$ of type \Hvier\  and
\item the non-degenerate biquadratics of $B$, $\bar{B}$, $C$ and $\bar{C}$ on edges not neighboring $A$ and $\bar{A}$.
\end{itemize}
Below is the list of all 3D consistent systems modulo $\left(\Moeb\right)^{8}$ with these properties and with the tetrahedron property. It turns out that in the last case the tetrahedron property follows from the above assumptions.
\begin{theo}  \label{Hsechs}Every 3D consistent system~(\ref{system}) satisfying the properties of this section and possessing the tetrahedron property is equivalent modulo $\left(\Moeb\right)^{8}$ to one of the following four systems which can be characterized by the quadruples of discriminants of $A$:
\begin{itemize}
\item $\left(0,0,0,0\right)$:
\begin{align*}
&A\left(x,x_{1},x_{2},x_{12}\right)=x+x_{1}+x_{2}+x_{12},\\
&B\left(x,x_{2},x_{3},x_{23};\alpha\right)=\left(x-x_{3}\right)\left(x_{2}-x_{23}\right)+\alpha,\end{align*}
\begin{align*}
 \bar{A}&=A\left(x_{3},x_{13},x_{23},x_{123}\right),&
\bar{B}&=B\left(x_{1},x_{12},x_{13},x_{123};\alpha\right),\\
 C&=B\left(x,x_{1},x_{3},x_{13};-\alpha\right),&
 \bar{C}&=B\left(x_{2},x_{12},x_{23},x_{123};-\alpha\right),\\
 K&=A\left(x,x_{12},x_{13},x_{23}\right),&
\bar{K}&=A\left(x_{1},x_{2},x_{3},x_{123}\right).\end{align*}
\item $\left(1,1,0,\delta_{1}\delta_{2}\right)$:
\begin{align*}
 &B\left(x,x_{2},x_{3},x_{23};\delta_{2}\right)=\left(x-x_{3}\right)\left(x_{2}-x_{23}\right)+\delta_{1}\alpha\left(x_{2}+x_{23}\right)+\delta_{2}\alpha\left(x+x_{3}\right)+\delta_{1}\delta_{2}\alpha^{2},\\
 \begin{split}\bar{B}\left(x_{1},x_{12},x_{13},x_{123};\delta_{1}\right)=\left(x_{1}-x_{13}\right)\left(x_{12}-x_{123}\right)+2\left(\delta_{1}-1\right)\alpha\\
+\left(\delta_{1}\delta_{2}+\delta_{1}-\delta_{2}\right)\alpha\left(x_{12}+x_{123}\right)+2\delta_{1}\delta_{2}\alpha x_{12}x_{123},\end{split}\\
&K\left(x,x_{13},x_{23},x_{12};\alpha\right)=x+x_{13}+\left(\delta_{1}-1\right)x_{23}+\delta_{1}\alpha+x_{12}\left(\delta_{2}x+\delta_{1}x_{23}+\delta_{1}\delta_{2}\alpha\right),\end{align*}
\begin{align*}
 A&=K\left(x,x_{1},x_{2},x_{12};0\right),&
\bar{A}&=K\left(x_{3},x_{13},x_{23},x_{123}\right),\\
 C&=B\left(x,x_{1},x_{3},x_{13};\delta_{1}+\delta_{2}-\delta_{1}\delta_{2}\right),&
\bar{C}&=\bar{B}\left(x_{2},x_{12},x_{23},x_{123};0\right),\\
 \bar{K}&=K\left(x_{3},x_{1},x_{2},x_{123};\alpha\right).\end{align*}
\begin{rem}
If $\delta_{1}=\delta_{2}=0$, $A$ is reducible.\end{rem}
\item $\left(x^{2},x_{1}^{2},x_{12}^{2},x_{2}^{2}\right)$:
\begin{itemize}
\item There are two non-equivalent systems with this quadruple of determinants:
\begin{align*}
 &A\left(x,x_{1},x_{2},x_{12};\delta_{1},\delta_{2}\right)=xx_{12}+x_{1}x_{2}+\delta_{1}x_{1}x_{12}+\delta_{2}x_{2}x_{12},\\
 &B\left(x,x_{2},x_{3},x_{23};\alpha\right)=xx_{23}+x_{2}x_{3}-\alpha\left(xx_{2}+x_{3}x_{23}\right)-\delta_{2}\left(\alpha^{2}-1\right)x_{2}x_{23},\end{align*}
\begin{align*}
 \bar{A}&=A\left(x_{3},x_{13},x_{23},x_{123};\delta_{1},\delta_{2}\right),&
\bar{B}&=B\left(x_{1},x_{12},x_{13},x_{123};\alpha\right),\\
 C&=B\left(x,x_{1},x_{3},x_{13};\alpha^{-1}\right),&
\bar{C}&=B\left(x_{2},x_{12},x_{23},x_{123};\alpha^{-1}\right),\\
 K&=A\left(x,x_{13},x_{23},x_{12};\delta_{1}\alpha^{-1},\delta_{2}\alpha\right),&
\bar{K}&=A\left(x_{3},x_{1},x_{2},x_{123};\delta_{1}\alpha^{-1},\delta_{2}\alpha\right)\end{align*}
\item and
\begin{align*}
 &A\left(x,x_{1},x_{2},x_{12};\delta_{1}\right)=xx_{2}+x_{1}x_{12}+\delta_{1}x_{1}x_{2}+\delta_{2}x_{2}x_{12},\\
 &B\left(x,x_{2},x_{3},x_{23};\delta_{1}\delta_{2}\right)=xx_{23}+x_{2}x_{3}-\alpha\left(xx_{2}+x_{3}x_{23}\right)+\delta_{1}\delta_{2}\left(\alpha^{2}-1\right)x_{2}x_{23},\end{align*}
\begin{align*}
 \bar{A}&=A\left(x_{3},x_{13},x_{23},x_{123};\delta_{1}\right),&
\bar{B}&=B\left(x_{123},x_{13},x_{12},x_{1};0\right),\\
 C&=B\left(x,x_{1},x_{3},x_{13};-\delta_{1}\right),&
\bar{C}&=B\left(x_{123},x_{23},x_{12},x_{2};-\delta_{1}\right),\\
 K&=A\left(x,x_{13},x_{23},x_{12};\delta_{1}\alpha\right),&
\bar{K}&=A\left(x_{3},x_{1},x_{2},x_{123};\delta_{1}\alpha\right).\end{align*}\end{itemize}\end{itemize}\end{theo}
\begin{rem}
These systems did not appear in the literature before. Special cases of equations $A=0$ in the last two systems are equivalent modulo $\left(\Moeb\right)^{4}$ to the equations whose integrability was shown in \cite{Hietarinta,LY} and the ``new'' integrable equation from \cite{HV}.\end{rem}

\subsection{Two Equations of Type \texorpdfstring{\Hvier}{H4}\  and Four Equations of Type \texorpdfstring{\Hsechs}{H6}}
In this section we consider systems~(\ref{system}) with
\begin{itemize}
\item $A$, $\bar{A}$, $C$ and $\bar{C}$ of type \Hsechs,
\item $B$ and $\bar{B}$ of type \Hvier\  and
\item the non-degenerate biquadratics of $B$ and $\bar{B}$ on diagonals.
\end{itemize}
Below is the list of all 3D consistent systems modulo $\left(\Moeb\right)^{8}$ with these properties and with the tetrahedron property.
\begin{theo}Every 3D consistent system~(\ref{system}) satisfying the properties of this section is equivalent modulo $\left(\Moeb\right)^{8}$ to one of the following six systems which can be characterized by the quadruples of discriminants of $A$:
\begin{itemize}
\item $\left(0,0,0,0\right)$:
\begin{align*}
 &A\left(x,x_{1},x_{2},x_{12}\right)=x+x_{1}+x_{2}+x_{12},\\
&B\left(x,x_{2},x_{3},x_{23};\alpha\right)=\left(x-x_{23}\right)\left(x_{2}-x_{3}\right)+\alpha,\end{align*}
\begin{align*}
 \bar{A}&=A\left(x_{3},x_{13},x_{23},x_{123}\right),&
\bar{B}&=B\left(x_{1},x_{12},x_{13},x_{123};\alpha\right),\\
 C&=A\left(x,x_{1},x_{3},x_{13}\right),&
\bar{C}&=A\left(x_{2},x_{12},x_{23},x_{123}\right),\\
 K&=B\left(x,x_{12},x_{13},x_{23};-\alpha\right),&
\bar{K}&=B\left(x_{1},x_{2},x_{3},x_{123};-\alpha\right).\end{align*}
\item $\left(1,1,0,\delta_{1}\delta_{2}\right)$:
\begin{itemize}
\item There are two non-equivalent systems with this quadruple of determinants:
\begin{align*}
 \begin{split}B\left(x,x_{2},x_{3},x_{23};\delta_{2}\right)=\left(x-x_{23}\right)\left(x_{2}-x_{3}\right)+\delta_{1}\alpha\left(x_{2}+x_{3}\right)+\delta_{2}\alpha\left(x+x_{23}\right)\\
 +\delta_{1}\delta_{2}\alpha^{2},\end{split}\\
 \begin{split}\bar{B}\left(x_{1},x_{12},x_{13},x_{123};\delta_{1}\right)=\left(x_{1}-x_{123}\right)\left(x_{12}-x_{13}\right)+2\left(\delta_{1}-1\right)\alpha\\
+\left(\delta_{1}\delta_{2}+\delta_{1}-\delta_{2}\right)\alpha\left(x_{1}+x_{123}\right)+2\delta_{1}\delta_{2}\alpha x_{1}x_{123},\end{split}\\
 &C\left(x,x_{1},x_{3},x_{13};\alpha\right)=x+x_{13}+\left(\delta_{1}-1\right)x_{3}+\delta_{1}\alpha+x_{1}\left(\delta_{2}x+\delta_{1}x_{3}+\delta_{1}\delta_{2}\alpha\right),\end{align*}
\begin{align*}
 A&=C\left(x,x_{1},x_{2},x_{12};0\right),&
\bar{A}&=C\left(x_{23},x_{123},x_{3},x_{13};0\right),\\
 \bar{C}&=C\left(x_{23},x_{123},x_{2},x_{12};\alpha\right),&
K&=B\left(x,x_{12},x_{13},x_{23};\delta_{1}+\delta_{2}-\delta_{1}\delta_{2}\right),\\
 \bar{K}&=\bar{B}\left(x_{1},x_{2},x_{3},x_{123};0\right)\end{align*}
\item and
\begin{align*}
&C\left(x,x_{1},x_{3},x_{13};\alpha\right)=x+x_{3}+\left(\delta_{1}-1\right)x_{13}+\delta_{1}\alpha+x_{1}\left(\delta_{2}x+\delta_{1}x_{13}+\delta_{1}\delta_{2}\alpha\right),\\
\begin{split}K\left(x,x_{12},x_{13},x_{23};\delta_{2}\right)=\left(x-x_{23}\right)\left(x_{12}-x_{13}\right)+\delta_{1}\alpha\left(x_{12}+x_{13}\right)\\
+\delta_{2}\alpha\left(x+x_{23}\right)+\delta_{1}\delta_{2}\alpha^{2},\end{split}\\
\begin{split}\bar{K}\left(x_{1},x_{2},x_{3},x_{123};\delta_{1}\right)=\left(x_{1}-x_{123}\right)\left(x_{2}-x_{3}\right)+2\left(\delta_{1}-1\right)\alpha\\
+\left(\delta_{1}\delta_{2}+\delta_{1}-\delta_{2}\right)\alpha\left(x_{1}+x_{123}\right)+2\delta_{1}\delta_{2}\alpha x_{1}x_{123},\end{split}\end{align*}
\begin{align*}
 A&=C\left(x,x_{1},x_{2},x_{12};0\right),&
\bar{A}&=C\left(x_{23},x_{123},x_{3},x_{13};0\right),\\
 B&=K\left(x,x_{2},x_{3},x_{23};\delta_{1}+\delta_{2}-\delta_{1}\delta_{2}\right),&
\bar{B}&=\bar{K}\left(x_{1},x_{12},x_{13},x_{123};0\right),\\
 \bar{C}&=C\left(x_{23},x_{123},x_{2},x_{12};\alpha\right).\end{align*}\end{itemize}
\begin{rem}
If $\delta_{1}=\delta_{2}=0$, $A$ is reducible.\end{rem}
\item $\left(x^{2},x_{1}^{2},x_{12}^{2},x_{2}^{2}\right)$:
\begin{itemize}
\item There are three non-equivalent systems with this quadruple of determinants:
\begin{align*}
 &A\left(x,x_{1},x_{2},x_{12};\delta_{1},\delta_{2}\right)=xx_{1}+x_{2}x_{12}+\delta_{1}x_{1}x_{12}+\delta_{2}x_{1}x_{2},\\
 &B\left(x,x_{2},x_{3},x_{23};\alpha\right)=xx_{3}+x_{2}x_{23}-\alpha\left(xx_{2}+x_{3}x_{23}\right)-\delta_{2}\left(\alpha^{2}-1\right)x_{2}x_{3},\end{align*}
\begin{align*}
 \bar{A}&=A\left(x_{23},x_{123},x_{3},x_{13};\delta_{1},\delta_{2}\right),&
\bar{B}&=B\left(x_{12},x_{1},x_{123},x_{13};\alpha\right),\\
 C&=A\left(x,x_{1},x_{3},x_{13};\delta_{1}\alpha^{-1},\delta_{2}\alpha\right),&
\bar{C}&=A\left(x_{23},x_{123},x_{2},x_{12};\delta_{1}\alpha^{-1},\delta_{2}\alpha\right),\\
 K&=B\left(x,x_{12},x_{13},x_{23};\alpha^{-1}\right),&
\bar{K}&=B\left(x_{2},x_{1},x_{123},x_{3};\alpha^{-1}\right),\end{align*}
\item moreover,
\begin{align*}
 &A\left(x,x_{1},x_{2},x_{12};\delta_{1}\right)=xx_{2}+x_{1}x_{12}+\delta_{1}x_{2}x_{12}+\delta_{2}x_{1}x_{2},\\
 &B\left(x,x_{2},x_{3},x_{23};\delta_{1}\delta_{2}\right)=xx_{3}+x_{2}x_{23}-\alpha\left(xx_{2}+x_{3}x_{23}\right)+\delta_{1}\delta_{2}\left(\alpha^{2}-1\right)x_{2}x_{3},\end{align*}
\begin{align*}
 &\bar{A}=A\left(x_{23},x_{123},x_{3},x_{13};\delta_{1}\right),&
&\bar{B}=B\left(x_{1},x_{12},x_{13},x_{123};0\right),\\
 &C=A\left(x,x_{1},x_{3},x_{13};\delta_{1}\alpha\right),&
&\bar{C}=A\left(x_{23},x_{123},x_{2},x_{12};\delta_{1}\alpha\right),\\
 &K=B\left(x,x_{12},x_{13},x_{23};-\delta_{1}\right),&
&\bar{K}=B\left(x_{1},x_{2},x_{3},x_{123};-\delta_{1}\right)\end{align*}
\item and last but not least
\begin{align*}
 &A\left(x,x_{1},x_{2},x_{12};\delta_{1}\right)=xx_{12}+x_{1}x_{2}+\delta_{1}x_{2}x_{12}+\delta_{2}x_{1}x_{12},\\
 &B\left(x,x_{2},x_{3},x_{23};-\delta_{1}\right)=xx_{3}+x_{2}x_{23}-\alpha\left(xx_{2}+x_{3}x_{23}\right)-\delta_{1}\left(\alpha^{2}-1\right)x_{2}x_{3},\end{align*}
\begin{align*}
 &\bar{A}=A\left(x_{23},x_{123},x_{3},x_{13};\delta_{1}\right),&
&\bar{B}=B\left(x_{1},x_{12},x_{13},x_{123};-\delta_{1}\right),\\
 &C=A\left(x,x_{1},x_{3},x_{13};\delta_{1}\alpha\right),&
&\bar{C}=A\left(x_{23},x_{123},x_{2},x_{12};\delta_{1}\alpha\right),\\
 &K=B\left(x,x_{12},x_{13},x_{23};\delta_{1}\delta_{2}\right),&
&\bar{K}=B\left(x_{1},x_{2},x_{3},x_{123};0\right).\end{align*}\end{itemize}\end{itemize}\end{theo}
\begin{rem}
Except two special cases (see \cite{Atk1}) these systems did not appear in the literature before.\end{rem}

\subsection{Six Equations of Type \texorpdfstring{\Hsechs}{H6}}
In this section we consider systems~(\ref{system}) with $A,\bar{A},\ldots,\bar{C}$ of type \Hsechs. Below is the list of all those systems possessing the tetrahedron property.
\begin{theo}Every 3D consistent system~(\ref{system}) satisfying the properties of this section is equivalent modulo $\left(\Moeb\right)^{8}$ to one of the following five systems which can be characterized by the quadruples of discriminants of $A$:
\begin{itemize}
\item $\left(0,0,0,0\right)$:
\[K\left(x,x_{12},x_{13},x_{23};\alpha,\beta\right)=\left(\alpha-\alpha\beta+\beta\right)x-\alpha\beta x_{12}-\alpha x_{13}+\beta x_{23},\]
\begin{align*}
 &A=K\left(x,x_{1},x_{2},x_{12};-1,1\right),&
&\bar{A}=K\left(x_{13},x_{3},x_{123},x_{23};\alpha\beta-\alpha-\beta,1\right),\\
 &B=K\left(x,x_{2},x_{3},x_{23};-\alpha,1\right),&
&\bar{B}=K\left(x_{12},x_{1},x_{123},x_{13};\left(\alpha\beta-\alpha-\beta\right)/\alpha,1\right),\\
 &C=K\left(x,x_{1},x_{13},x_{3};1,-\beta\right),&
&\bar{C}=K\left(x_{12},x_{2},x_{23},x_{123};1,\left(\alpha\beta-\alpha-\beta\right)/\beta\right),\\
 &\bar{K}=K\left(x_{123},x_{3},x_{2},x_{1};\alpha,\beta\right).\end{align*}
\item $\left(0,0,1,1\right)$:
\begin{itemize}
\item There are two non-equivalent systems with this quadruple of determinants:
\begin{align*}
&A\left(x,x_{1},x_{2},x_{12}\right)=xx_{1}-x_{2}-x_{12},&
&\bar{A}\left(x_{3},x_{13},x_{23},x_{123}\right)=\left(x_{3}+x_{13}\right)x_{23}+x_{123},\end{align*}
\begin{align*}
&B=A\left(x,x_{23},x_{2},x_{3}\right),&
&\bar{B}=\bar{A}\left(x_{12},x_{13},x_{1},x_{123}\right),&
&C=A\left(-x,x_{1},x_{3},x_{13}\right),\\
&\bar{C}=\bar{A}\left(x_{2},x_{12},x_{23},-x_{123}\right),&
&K=A\left(-x,x_{23},x_{12},x_{13}\right),&
&\bar{K}=\bar{A}\left(x_{2},x_{3},x_{1},-x_{123}\right)\end{align*}
\item and
\begin{align*}
 &A\left(x,x_{1},x_{2},x_{12}\right)=xx_{1}-x_{2}-x_{12},\\
&\bar{A}\left(x_{3},x_{13},x_{23},x_{123}\right)=\left(x_{3}+x_{13}\right)x_{123}+x_{3}-x_{23},\end{align*}
\begin{align*}
 &B=\bar{A}\left(x_{2},x_{3},-x_{23},\alpha^{-1}x\right),&
&\bar{B}=A\left(\alpha x_{1},x_{123}+1,-x_{12},-x_{13}\right),\\
 &C=A\left(x+\alpha,x_{1},-x_{3},-x_{13}\right),&
&\bar{C}=\bar{A}\left(x_{2},x_{12},-x_{23},x_{123}\right),\\
 &K=\bar{A}\left(x_{12},x_{13},x_{23},\alpha^{-1}x\right)&
&\bar{K}=A\left(\alpha x_{1},x_{123},x_{2},x_{3}\right).\end{align*}\end{itemize}
\item $\left(x^{2},x_{1}^{2},x_{12}^{2},x_{2}^{2}\right)$:
\begin{itemize}
\item There are two non-equivalent systems with this quadruple of determinants:
\[ A\left(x,x_{1},x_{2},x_{12};\delta_{1},\delta_{2}\right)=xx_{12}+x_{1}x_{2}+\delta_{1}x_{1}x_{12}+\delta_{2}x_{2}x_{12},\]
\begin{align*}
 &\bar{A}=A\left(x_{13},x_{3},x_{123},x_{23};\delta_{1},\delta_{2}\right),&
&B=A\left(x,x_{2},x_{3},x_{23};\delta_{2},0\right),\\
 &\bar{B}=A\left(x_{13},x_{1},x_{123},x_{12};\delta_{1}\delta_{3},\delta_{2}\right),&
&C=A\left(x_{13},x_{3},x,x_{1},\delta_{1},-\delta_{3}\right),\\
 &\bar{C}=A\left(x_{123},x_{23},x_{2},x_{12},\delta_{1},-\delta_{3}\right),&
&K=A\left(x_{13},x,x_{23},x_{12};-\delta_{3},-\delta_{1}\delta_{2}\right),\\
 &\bar{K}=A\left(x_{3},x_{1},x_{123},x_{2};-\delta_{3},0\right)&\end{align*}
\item and
\[ A\left(x,x_{1},x_{2},x_{12};\delta_{1},\delta_{2}\right)=xx_{1}+x_{2}x_{12}+\delta_{1}x_{1}x_{12}+\delta_{2}x_{1}x_{2},\]
\begin{align*}
 &\bar{A}=A\left(x_{13},x_{3},x_{123},x_{23};\delta_{1},\delta_{2}\right),&
&B=A\left(x,x_{3},x_{2},x_{23};\delta_{2},0\right),\\
 &\bar{B}=A\left(x_{13},x_{1},x_{123},x_{12};\delta_{1}\delta_{3},\delta_{2}\right),&
&C=A\left(x_{12},x_{1},x_{3},x;-\delta_{3},-\delta_{1}\delta_{2}\right),\\
 &\bar{C}=A\left(x_{123},x_{12},x_{23},x_{2};-\delta_{3},0\right),&
&K=A\left(x_{13},x_{12},x,x_{23};\delta_{1},-\delta_{3}\right),\\
 &\bar{K}=A\left(x_{123},x_{1},x_{2},x_{3};\delta_{1},-\delta_{3}\right).\end{align*} \end{itemize}\end{itemize}\end{theo}
\begin{rem}
All systems except for the one characterized by the quadruple $\left(0,0,0,0\right)$ did not appear in the literature before. The latter is equivalent to a special case of the one of the systems from \cite{Atk2} (in that paper the tetrahedron property is not assumed).\end{rem}

\section{Embedding in the lattice \texorpdfstring{$\Z^{3}$}{Z\Dach3}} \label{embedding}
The main conceptual message of \cite{quadgraphs,ABS1} is that 3D consistency is synonymous with integrability. In the situations considered there, where equations on opposite faces of the cube are shifted versions of one another, it was demonstrated how to derive B\"acklund transformations and zero curvature representations from a 3D consistent system.\par
It might be not immediately obvious whether these integrability attributes can still be derived for our systems, where the equations on opposite faces of one elementary cube happen to be completely different. We show now this is the case, indeed.\par
\begin{figure}[htbp]
   \centering
   \includegraphics{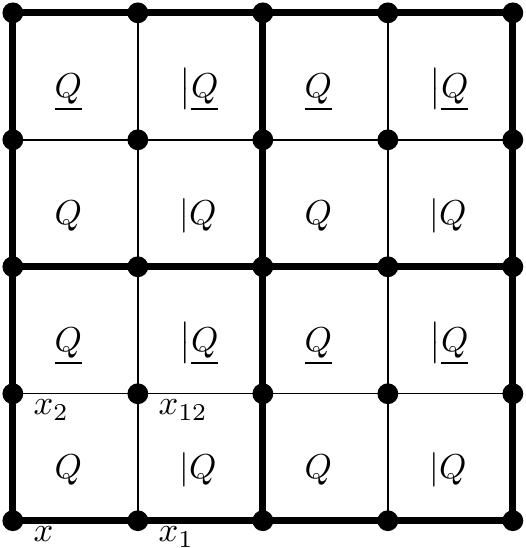} 
   \caption{Embedding in a planar lattice}
   \label{fig:flat}
\end{figure}
We start with the composing an integrable system on $\Z^{2}$ from a non-symmetric multi-affine polynomial $Q$ from one of our lists. The polynomials will be assigned to the faces as demonstrated in Figure~\ref{fig:flat}. For an equation
\[Q\left(x,x_{1},x_{2},x_{12}\right)=0\]
we define
\begin{align*}&\left|Q\right.:=Q\left(x_{1},x,x_{12},x_{2}\right),&
&\underline{Q}:=Q\left(x_{2},x_{12},x,x_{1}\right)\quad \text{and}&
&\left|\underline{Q}\right.:=Q\left(x_{12},x_{2},x_{1},x\right).\end{align*}
This can be interpreted as reflections at the axis implied by the notation. So, the basic elements of our embedding are not as usual faces but quadruples of faces as marked by the bold lines in Figure~\ref{fig:flat} and the embedding is not one-periodic as usual but two-periodic in each direction. In the cases considered in \cite{XP} the equation $\left|\underline{Q}\right.$ is a shifted version of $Q$ as well as $\underline{Q}$ is a shifted version of $\left|Q\right.$.\par
We show now that this 2D system, with an elementary $2\times 2$ building-block, is integrable: One can find (properly generalized) B\"acklund transformations and zero curvature representations for these systems.\par
\begin{figure}[htbp]
   \centering
   \subfloat[Symmetric case]{\label{fig:cube3}\includegraphics{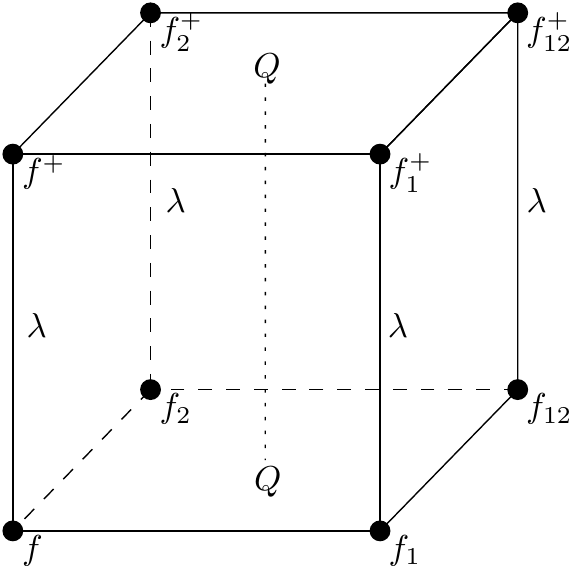}}\qquad
   \subfloat[Non-symmetric case]{\label{fig:cube4}\includegraphics{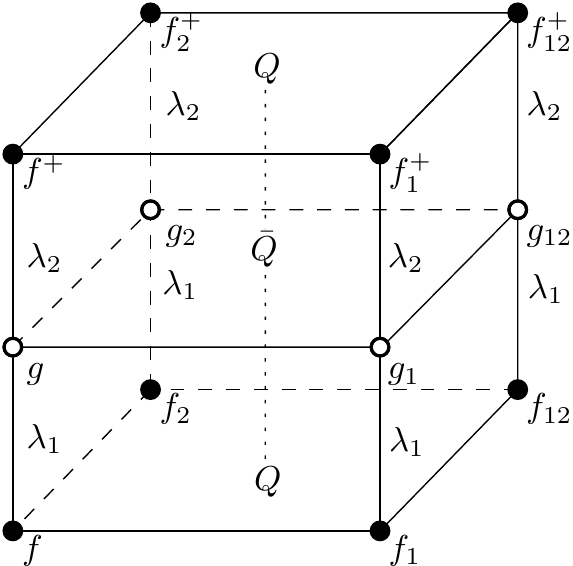}}
   \caption{B\"acklund transformation}
\end{figure}
Let us start with B\"acklund transformations. In the symmetric case, i.e. for systems of the ABS-list, we have the picture like in Figure~\ref{fig:cube3}. A B\"acklund transformation can be interpreted as one layer of the system in the three dimensional lattice. We have a solution $f:V\left(\D\right)\to\C$ on a quad-graph $\D$ of 
\[Q\left(f,f_{1},f_{2},f_{12}\right)=0\]
on the ground floor, its B\"acklund transformation $f^{+}:V\left(\D^{+}\right)\to\C$ on a copy $\D^{+}$ of $\D$ with 
\[Q\left(f^{+},f_{1}^{+},f_{2}^{+},f_{12}^{+}\right)=0\]
on the first floor and the B\"acklund parameter assigned to the vertical edges. A more detailed demonstration of this situation can be found for example in \cite{ddg}. In the non-symmetric case, i.e. for our systems, we have to consider a picture which is a little bit more extensive, as demonstrated in Figure~\ref{fig:cube4}. In this case a B\"acklund transformation can be seen as two layers of our lattice. We start again with a solution $f:V\left(\D\right)\to\C$ of
\[Q\left(f,f_{1},f_{2},f_{12}\right)=0\]
on the ground floor and get a transformation $g:V\left(\bar{\D}\right)\to\C$ on a copy $\bar{\D}$ of $\D$ with the equation on the opposite face of the cube
\[\bar{Q}\left(g,g_{1},g_{2},g_{12}\right)=0\]
and a parameter $\lambda_{1}$ assigned to the vertical edges. Every parameter of the system we consider which do not appear in $Q$ and $\bar{Q}$ can be chosen as $\lambda_{1}$. Then, starting from $g$ we get a B\"acklund transformation of $f$ called $f^{+}:V\left(\D\right)\to\C$ on the second floor with
\[Q\left(f^{+},f_{1}^{+},f_{2}^{+},f_{12}^{+}\right)=0\]
and a parameter $\lambda_{2}$ assigned to the vertical edges.\par
\begin{figure}[htbp]
   \centering
   \subfloat[Symmetric case]{\label{fig:cube5}\includegraphics{figure_2a}}\qquad
   \subfloat[Non-symmetric case]{\label{fig:cube6}\includegraphics{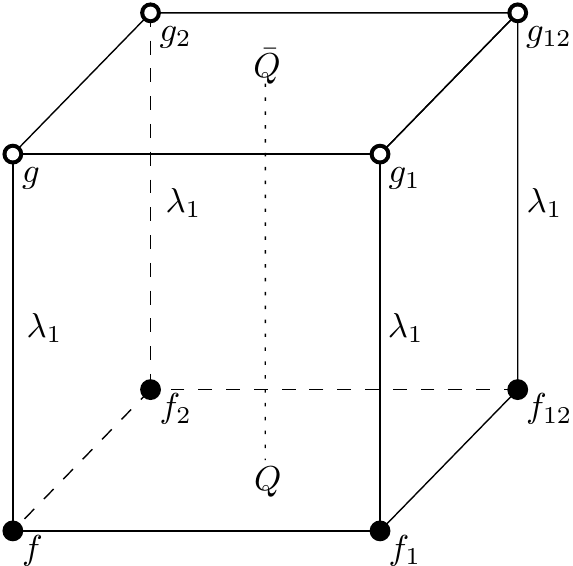}}
   \caption{Zero curvature represenation}
\end{figure}
Zero curvature representations can be derived for non-symmetric systems, too. We will first consider briefly the idea how to derive zero curvature representations of the symmetric systems. In the symmetric case we have again the picture like in Figure~\ref{fig:cube5}. In this case a transition matrix of a zero curvature representation of an equation on the ground floor can be interpreted as a M\"obius transformation (in the standard matrix notation) from one vertex of the first floor to another one connected by an edge e.g.,
\[f_{1}^{+}=L\left(f,f_{1},\lambda\right)\left[f^{+}\right]\]
with $L$ a $2\times 2$ transition matrix dependent on the spectral parameter $\lambda$. One can derive the M\"obius transformation from the equation of the corresponding face. For informations in more details we again refer to \cite{ddg}. In the non-symmetric case we also need just one layer to derive a zero curvature equation (see Figure~\ref{fig:cube6}). Also in this case a transition matrix of a zero curvature representation of an equation on the ground floor can be interpreted as a M\"obius transformation from one vertex of the first floor to another one conneted by an edge in the standard matrix notation, e.g.,
\[g_{1}=L\left(f,f_{1},\lambda_{1}\right)\left[g\right]\]
with $L$ a $2\times 2$ transition matrix dependent on the spectral parameter $\lambda_{1}$.

\section{Concluding Remarks}
Due to Section~\ref{embedding} the Theorems~\ref{Hvierfirst}, \ref{Hviersecond} and \ref{Hsechs} gives us a B\"acklund transformation and a zero curvature representation for every quad-equation of type~\Hvier\  and of type~\Hsechs\ in every arrangement of fields to the vertices of an elementary square of $\Z^{2}$. Therefore, they are all integrable.\par
Moreover, our classification includes all known 3D consistent systems except two systems mentioned in \cite{Atk1}, the system mentioned in \cite{Hietarinta} and two systems only containing linear equations (see \cite{Atk2,ABS2}). Of course, all these system do not possess the tetrahedron property.

\section*{Acknowledgments}
The author is supported by the Berlin Mathematical School and is indebted to Yuri~B. Suris for his continued guidance.

\bibliographystyle{amsalpha}
\bibliography{Quellen}

\end{document}